\documentclass[fleqn,floatfix,usenatbib]{mnras}

\usepackage{newtxtext,newtxmath}
\usepackage{comment}
\usepackage{adjustbox}
\usepackage{longtable}

\usepackage[T1]{fontenc}

\DeclareRobustCommand{\VAN}[3]{#2}
\let\VANthebibliography\thebibliography
\def\thebibliography{\DeclareRobustCommand{\VAN}[3]{##3}\VANthebibliography}

\usepackage{graphicx}	
\usepackage{amsmath}	
\usepackage{lipsum}
\usepackage{orcidlink}

\title[Tidal dwarf galaxies]{Structural and dynamical properties of Tidal dwarf galaxies in the tails and bridge of the Guitar galaxy Arp 105}

\author[Prakash \& Saha]{
Jyoti Prakash$^{1}$\thanks{E-mail: jyoti.prakash@iucaa.in}\orcidlink{0009-0006-6208-8863},
Kanak Saha$^{1}$\orcidlink{0000-0002-8768-9298}
\\
$^{1}$ Inter-University Centre for Astronomy and Astrophysics, Pune 411007, India.\\
}

\date{}

\begin{document}
\maketitle

\begin{abstract}
We present a multi-wavelength (far-ultraviolet to infrared) analysis of two tidal dwarf galaxy (TDG) candidates and a tidal bridge in the interacting system Arp 105 at $z = 0.029$ in the Abell 1185 cluster. Far-ultraviolet observations obtained with the Ultraviolet Imaging Telescope onboard AstroSat reveal strong FUV emission from the tidal galaxies Arp~105N and Arp~105S, indicating recent star formation. In Arp~105N, strong nebular emission lines and large equivalent widths $EW(H\alpha)=77.6\pm1.3$~\AA\ and $EW(H\beta)=15.8\pm1.7$~\AA\, imply a dominant starburst age of $\sim$6 - 10~Myr under an instantaneous-burst assumption, while the FUV emission suggests star formation sustained over the past $\sim100 - 200$~Myr. The relatively high metallicity, $\sim2/3,Z_\odot$ (based on the strong-line method), is consistent with expectations for a tidal dwarf galaxy formed from material inherited from the host galaxy. Together, these results suggest that Arp~105N hosts a composite stellar population, consisting of older stars stripped from the host galaxy and younger stars formed in situ. Spectral energy distribution modeling yields stellar masses of $5.75\times10^{9}$, $0.8\times10^{9}$, and $6.8\times10^{9},\rm M_\odot$ for Arp~105N, Arp~105S, and the tidal bridge, respectively. Based on the dynamical mass estimate from VLA HI measurements for Arp~105N and based on CFHT H$\alpha$ kinematics for A105S, they have a dynamical-to-baryonic mass ratio of $\sim1.95$ and $\sim1.30$, respectively, indicating a deficiency of dark matter. Further observations, particularly integral field spectroscopy and high-resolution 21~cm observations, may provide better constraints into the kinematics and improve understanding of TDG formation.
\end{abstract}

\begin{keywords}
galaxies: interactions-galaxies: dwarf-galaxies: fundamental parameters.
\end{keywords}



\section{Introduction}
\label{sec:Intro}
In the standard model of hierarchical structure formation, galaxies are thought to have assembled via the successive merging and accretion of smaller systems \citep{r105, Fall1983, LHuillieretal2012}. Within this framework, \textit{dwarf galaxies} — defined by their relatively low stellar masses ($M_{*} \sim 10^7$--$10^{9.5}~\rm M_\odot$), relatively smaller optical sizes (typically 0.1--10\,kpc), and low metallicities — are considered the most numerous and fundamental building blocks of more massive galaxies \citep{r43,r44,r72, Reines2013,r73}. The galaxy mass function observed in the local Universe reveals that dwarf galaxies are the most abundant type of galaxies \citep{Schechter1976,sandage1985, van1992, Sabatini, Kelvin2012, Moustakas2013, Wright2017, kadofong2025}.

Classical dwarf galaxies such as dwarf irregulars (dIrr), dwarf spheroidals (dSph), and blue compact dwarfs (BCDs) are typically dark matter \citep{r1} dominated and exhibit a wide variety of star formation histories, morphologies, and internal kinematics \citep{r74,r99,r75}. Both observations and simulations confirm their dominant contribution to the low-mass end of the galaxy luminosity and stellar mass functions in the local Universe \citep{r103,r104}. Dwarf galaxies fall in a crucial regime for testing both galaxy formation physics and the nature of dark matter.

However, a growing body of observational and theoretical work has revealed an alternative formation pathway for some dwarf galaxies — one that does not involve the collapse of primordial fluctuations. In that, \textit{tidal dwarf galaxies (TDGs)} or \textit{non-classical dwarf galaxies} are formed during binary interactions or mergers between gas-rich disk galaxies or between a spiral and an elliptical. In such encounters, tidal forces can expel substantial amounts of interstellar gas and stars manifested in the form of long tidal tails and bridges around them \citep{r60, Wright1972,r83,Yun, Martinetal2022, Namumba, Barnesetal1992,Mihos_2001, mihos2004}. Under favorable conditions, self-gravitating condensations can form in these structures, collapsing to create dwarf-galaxy-sized systems \citep{r3,r13,r14,r15,r16, DumontAmelie2021}. The formation rate of TDGs is thought to have been higher at earlier cosmic times, driven by increased gas fractions and enhanced frequency of interactions. However, the survival fraction of TDGs remains uncertain\citep{kroupa1997, Okazaki_2000,r15, Kaviraj, Dabri_nd_Kroupa}. Numerical simulation shows that these TDGs can breathe up to many Gyrs \citep{kroupa1997, Recchi2007, Klessen_1998, Casas2012, Ploeckinger2014, Ploeckinger2015}. Theoretical work done by  \cite{Bournaud2010} suggests that $\sim50\%$ of TDGs with stellar masses $M_* > 10^8\rm M_\odot$ have a higher chance to survive up to Hubble time.

Unlike their classical counterparts, TDGs are composed of pre-enriched, recycled material, and are therefore expected to be metal-rich for their mass, occupying a region above the Mass-Metallicity relation \citep{r3, Weilbacher2003, Recchi2015}. TDGs are thought to contain little or no dark matter \citep{r17,r53,r54,r55}. Simulations predict that TDGs can form in mergers with mass ratios greater than 1:8, and that they may account for up to $\sim$30\% of the dwarf galaxy population in certain environments \citep{r42}. Observational surveys in the HI 21\, cm line suggest that nearly 25\% of nearby galaxies exhibit signs of recent or ongoing tidal interactions, indicating that the formation of TDGs may not be a rare phenomenon \citep{r9}. The HI component of gas-rich galaxies extends beyond the optical disk, making it more vulnerable to tidal stripping \citep{Hibbard_2000,koiri2018,kori2024, Martin}. Despite their importance, the identification and confirmation of TDGs remain observationally challenging. Their dynamical and chemical properties must be carefully disentangled from classical dwarfs, and their formation history, metallicity, kinematics, and stellar populations offer critical diagnostics for doing so. The study of TDGs therefore holds an important clue to understanding galaxy evolution, the baryon cycle, and even tests of alternative gravity theories \citep{r6,r7}.

In this context, we present a detailed multi-wavelength investigation of the interacting galaxy system Arp\,105, also known as the `\textit{`Guitar Galaxy}'' due to its distinctive morphology. \text{Arp\,105} is an ongoing merger between a luminous spiral galaxy (NGC\,3561A; $z_{\rm spec} = 0.030$) and an elliptical companion (NGC\,3561B; $z_{spec} = 0.029$), located in the Abell 1185 galaxy cluster in the constellation \textit{Ursa Major} \citep{r2}. First noted by \citet{r82}, this system exhibits extended tidal structures, including long H{\sc i}-rich tails and bridges. Follow-up observations by \citet{r60} and \citet{r83} identified blue stellar condensations along these tidal features, which have since been interpreted as tidal dwarf galaxy candidates. In particular, two prominent TDG candidates named as Arp\,105N\,(A105N):(RA = 11:11:12.8, Dec = +28:45:57.1) and Arp\,105S\,(A105S):(RA = 11:11:13.4, Dec = +28:41:15.9) — lie at the ends of the northern and southern tidal extensions.\cite{r2} described A105N as a Magellanic Irregular type galaxy and A105S as a blue compact galaxy. Based on VLA mapping of HI and CO(1–0) emission, \cite{r50} analyzed gas segregation in the interacting system Arp 105. Absorption lines reveal that the interaction between NGC 3561A and B directly drives mass transfer from a spiral to an elliptical galaxy. Velocity discrepancies in the southern HI clouds (associated with NGC 3561B and A105S) suggest distinct origins or the partial collapse of a single structure that formed the tidal dwarf A105S. Similarly, a kinematically decoupled component within the northern HI tail is associated with the star-forming dwarf A105N. Both tidal objects exhibit possible signs of rotation. \cite{Smith2010} suggests, based on analogy to Arp 285 \citep{Smith2008} and proximity to the elliptical, the southern star-forming region in Arp 105 is an accretion tail, rather than simply a classical tidal tail coincidentally seen in projection behind the elliptical galaxy.

These objects present an ideal laboratory for studying TDG formation, evolution, and the effect of environment, as they lie in a dynamically active system with clearly observable tidal debris. 
In this work, we perform a comprehensive analysis of Arp\,105 and its associated tidal features, including both TDG candidates and the connecting stellar bridge. Using multi-wavelength data spanning the far-ultraviolet (FUV) to the infrared (IR), we derive star formation rates, stellar ages, colors, and structural properties of these regions. This analysis provides new constraints on the physical processes governing the formation and survivability of TDGs, and fills a gap in the current literature by presenting one of the few spatially resolved studies of both the bridge and TDG components in a single interacting system.  A distinguishing factor of this study is the inclusion of AstroSat FUV data, which offers superior spatial resolution (PSF $\approx 1.4''$) compared to previous studies that relied solely on GALEX NUV photometry (PSF $\approx 5''$) by \cite{Smith2010}.

Throughout the paper, we adopt a standard cosmology with $\Omega_{\rm m}=0.3, \Omega_{\rm \Lambda}=0.7$ and use $H_0 = 70\, \rm{km s^{-1} Mpc^{-1}}$. Our physical scale size is 0.581 kpc/arcsec using the cosmology calculator \citep{r86}. The luminosity distance $D_L$ is 127.0 Mpc. All magnitudes are given in the AB system \citep{AB_mag}. Throughout this study, we assumed the Salpeter Initial Mass Function (IMF) \citep{r69}, unless stated otherwise. We use shorthand notations A105N(N1, N2) and A105S for Arp 105N and Arp 105S, respectively.

This paper is structured as follows. In Section 2, we provide details of the observational data used in this study. Section 3 describes the photometry method: background subtraction, 2D PSF modeling and matching, source extraction, flux measurement, and extinction correction methods. In Section 4, we discussed spectral flux measurement, correction, and metallicity. In Section 5, we present the stellar mass estimates using \texttt{CIGALE} modeling and discuss the color–mass relation. Section 6 focuses on spectral analysis and discusses the star formation rate. Finally, Section 7 follows discussions, conclusions, and a summary are presented.  Section 8, followed by supplementary materials in the appendix.

\section{AstroSat observation and archival data on Arp 105}
\label{sec:obs}

\subsection{AstroSat Observation}
The interacting galaxy system Arp 105 was observed (Proposal ID: A04\_201; PI: Kanak Saha) for 6000 sec (on target) on May 29, 2018, in photon counting mode by the Ultra-Violet Imaging Telescope (UVIT) on AstroSat. We obtained the raw dataset as Level 1 (L1) FITS files from the AstroSat archive  \footnote{\url{https://astrobrowse.issdc.gov.in/astro-archive/archive/Home.jsp}} \citep{Balamuruganetal2021}. The L1 dataset for each orbit contains the science data from two filters i.e., F154W ($\sim 1300 - 1800$\AA), VIS2 channel, as well as the Auxiliary data, which provides critical information such as the time calibration table, orbital position, attitude information, and housekeeping information of the satellite during the entire period of observation. We processed the L1 data to create Level 2 (L2) images using the official UVIT L2 pipeline (hereafter UL2P, \citep{GHOSH_2022}). Each orbit data (i.e., the L2 data) was then co-added (weighted by the exposure map for each orbit, see \cite{Sahaetal2024} for details) to create a science-ready image. The exposure time for the final science-ready image is 5426 sec. It is to be noted that there was no GALEX tile for the FUV observation for this galaxy field, as well as for any bright star. For the astrometric alignment, we utilized 8 point-like sources spread across the full FoV (28' diameter) by cross-identifying them from the GALEX NUV and SDSS g-band, i.e., we extracted the x pixel/y pixel from the F154W image and their corresponding RA/Dec from the GALEX field. We use an IDL program that takes these inputs and performs a TANGENT-Plane astrometric plate solution similar to ccmap task of IRAF\footnote{\url{https://iraf-community.github.io/}}. Apart from star-tracking, we do not use the VIS2 image for any photometric measurement. The final astrometric accuracy in the F154W filter image has an RMS $\sim 0.36"$ (note, a pixel $\simeq 0.417"$ in the UVIT sub-pixel image). The photometric calibration is performed with a white dwarf star Hz4; the updated photometric zeropoint is 17.778 AB mag for the F154W filter \citep{r25}. 

\subsection{Ancillary data: SDSS and Spitzer}
 In addition to FUV, we have utilized archival imaging data from SDSS\footnote{\url{https://skyserver.sdss.org/dr16/en/tools/chart/navi.aspx}} (Sloan Digital Sky Survey) DR16 \citep{r27} in u/g/r/i/z bands as well as archival imaging data from the Spitzer Space Telescope in this study. The Infrared Array Camera (IRAC1/2/3/4)\footnote{\url{https://sky.esa.int/}} captures images at four filters centered at 3.6, 4.5, 5.8, and 8.0\,$\mu$m \citep{r28}. Of the three tidal objects of interest (A105N, A105S, and tidal bridge), only A105N has spectroscopic observation\footnote{\url{https://dr16.sdss.org/sas/dr16/sdss/spectro/redux/26/spectra/lite/2213/spec-2213-53792-0171.fits}} obtained using the SDSS Baryon Oscillation Spectroscopic Survey (BOSS) instrument. As A105N contains 2 knots, the spectra are centered around the coordinate RA: 167.80288 and Dec: 28.76550 deg, and fiber diameter 3 arcsecs, which corresponds to A105N2 (see Figure.\ref{seg}).

\begin{figure}
     \centering
     \includegraphics[width=1.1\linewidth]{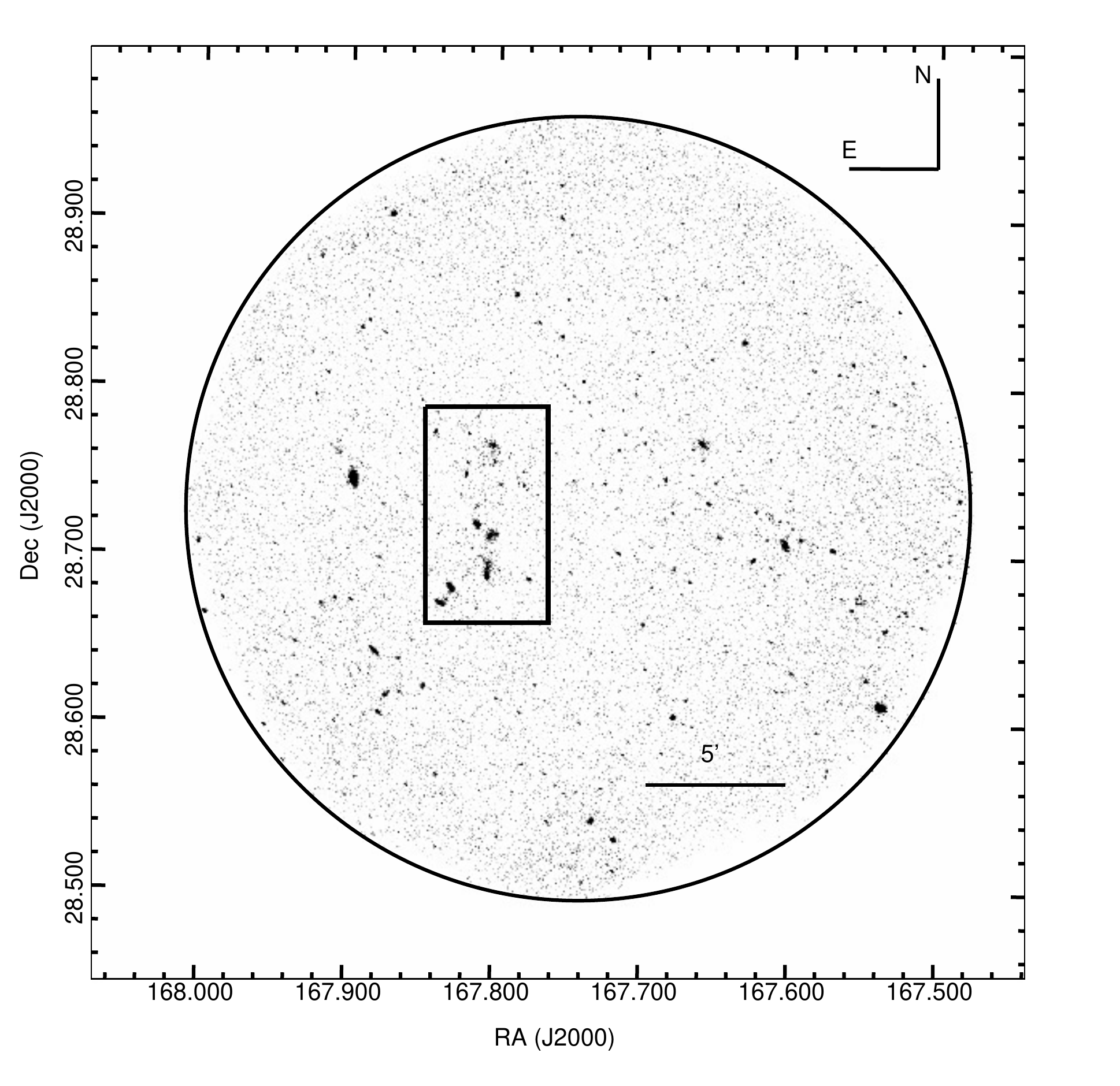}
     \caption{F154W (FUV) band image with a full field of view (shown as dark solid circle of diameter 28 arcminutes). The black rectangular box region centered at Arp 105 is an interacting galaxy system. North direction is up. }
     \label{fig:enter-label}
 \end{figure}
 
\begin{figure}
\centering
\includegraphics[width=1.\linewidth,angle=0]{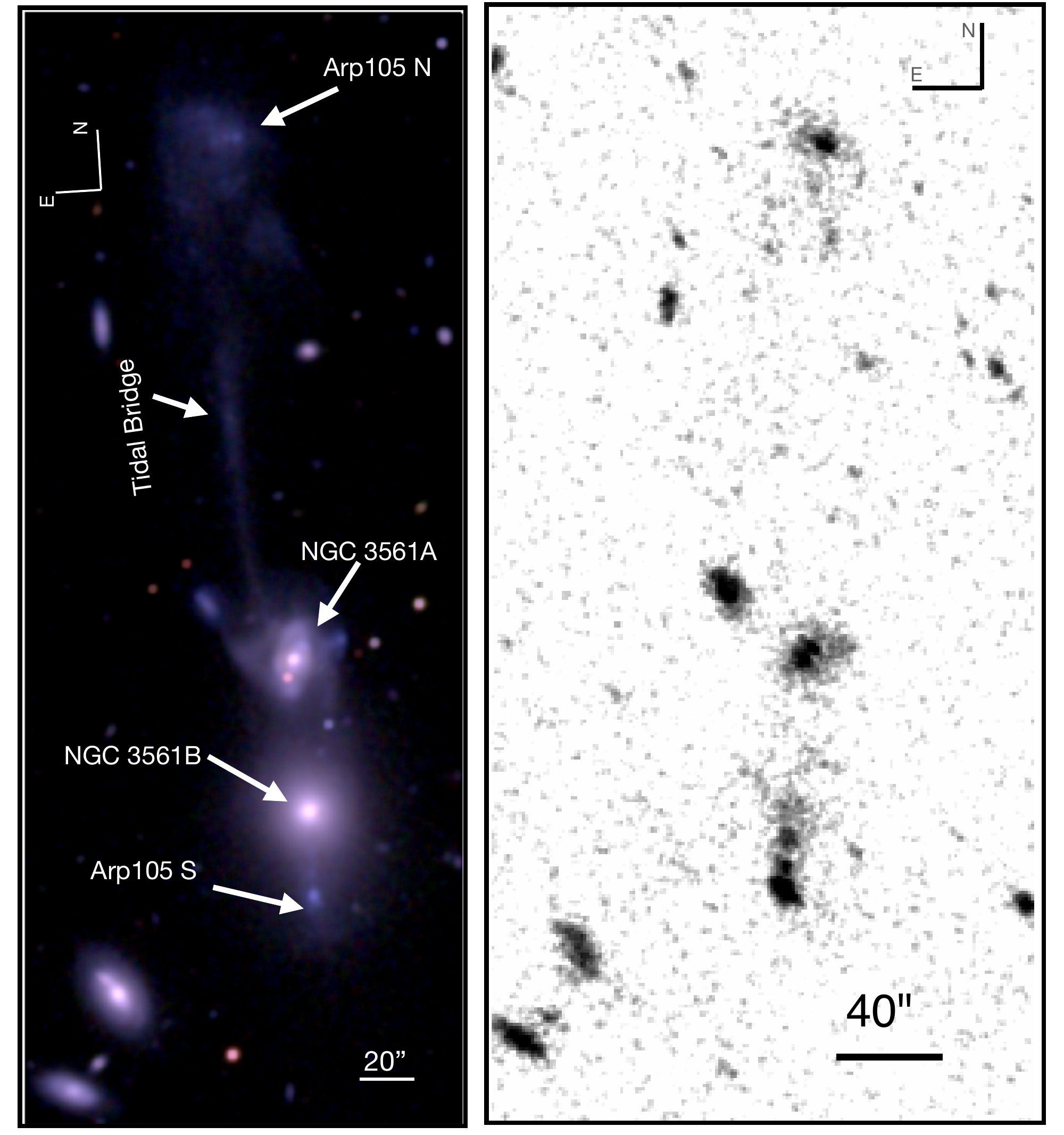}
\caption{(Left panel): SDSS false-color image of the Arp 105 interacting system created using gri bands. The tidal bridge connects the tidal dwarf galaxy A105N, located at the extreme North, and the starburst galaxy NGC 3561A. (Right panel): AstroSat FUV image in grayscale, inverted. A tophat smoothing kernel with a 3-pixel radius was applied to both images. North direction is up}
\label{RGB}
\end{figure}

\section{Data analysis: Surface Photometry} 
\label{sec:dataanalysis}
This section covers background measurement, source extraction, PSF modeling, 2D \footnote{\url{https://users.obs.carnegiescience.edu/peng/work/galfit/galfit.html}}\texttt{GALFIT} modeling, flux measurement, and corrections.

\subsection{Background, PSF modeling and PSF-matching}
\label{sec:PSF}

\begin{figure}
        \centering
        \includegraphics[width=1.0\linewidth]{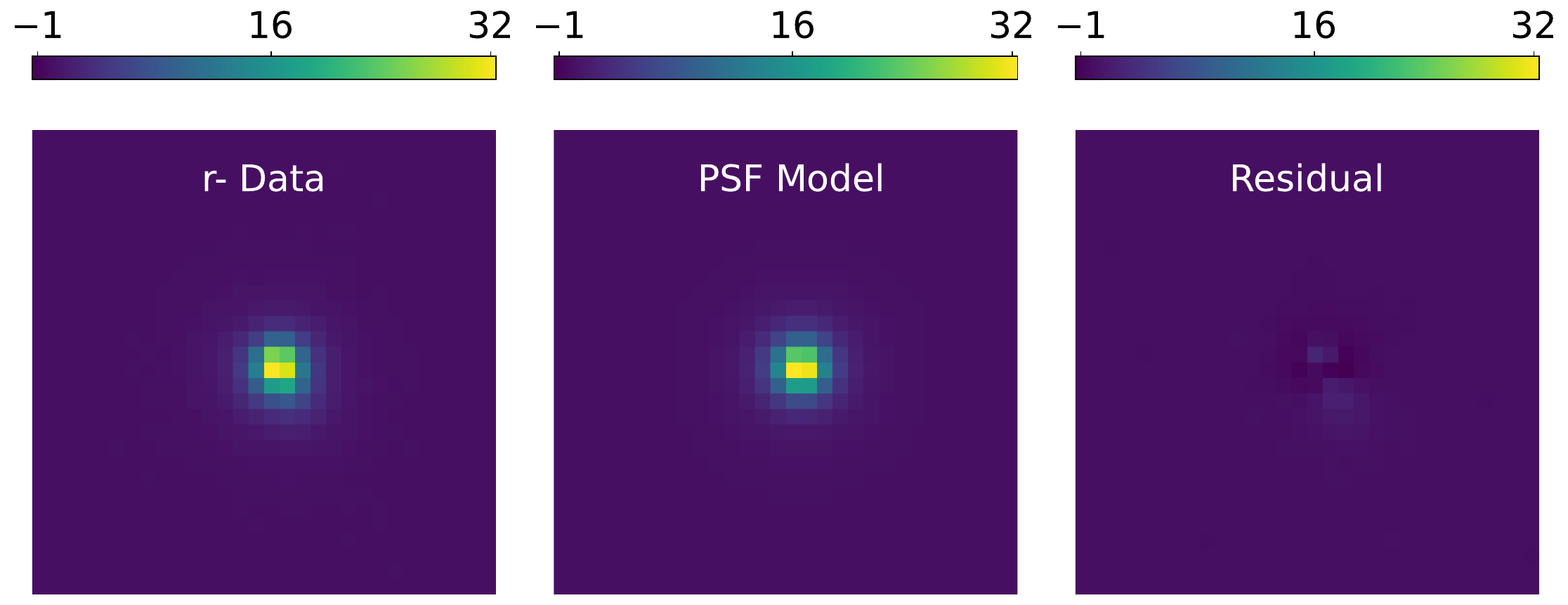}
        \includegraphics[width=1.0\linewidth]{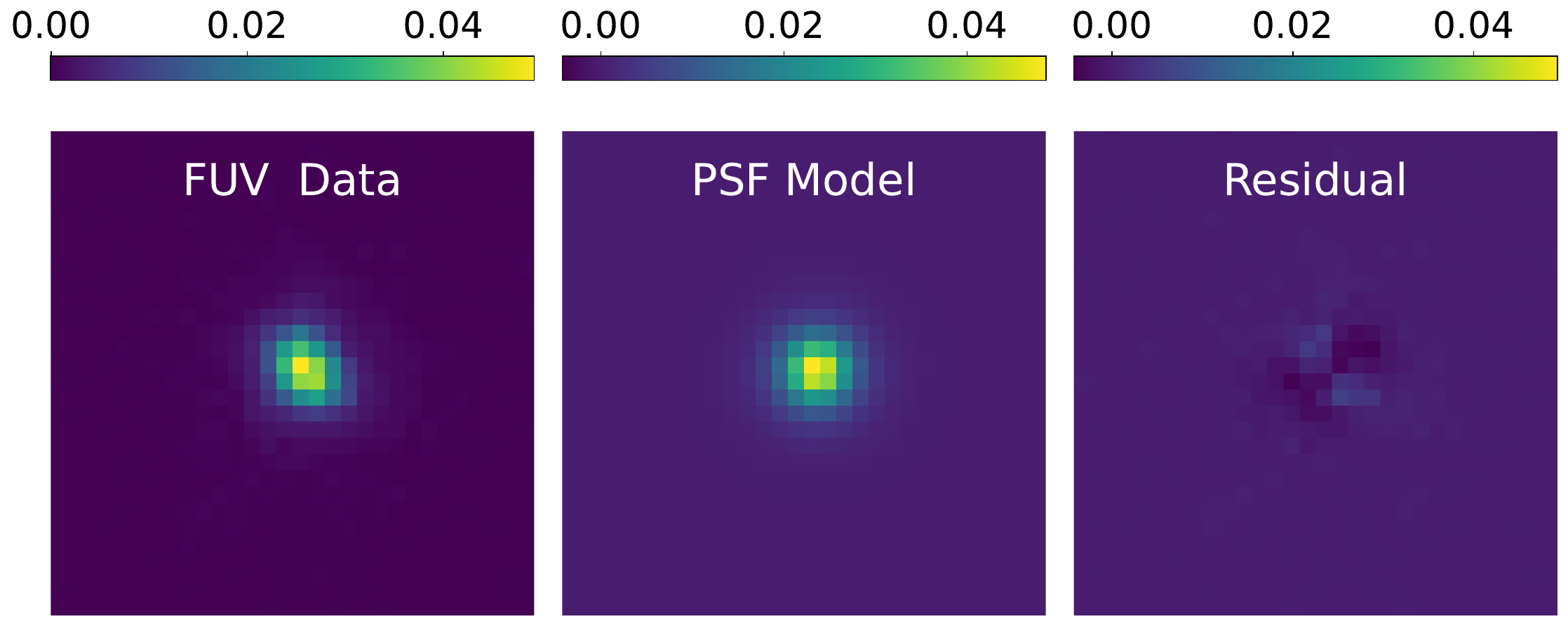}
        \caption{Two-dimensional Moffat profile fitting of a stellar point spread function (PSF) for the SDSS r band (top row) and AstroSat FUV band (bottom row). From left to right: the original observed data, the best-fit 2D model, and the resulting residuals ($\text{Data} - \text{Model}$). The color bar units are nanomaggies (NMgy) and count/s, respectively. Their image cutout size is 31$\times$31 pixels.}
       \label{psf}
\end{figure}

We estimate the background for the F154W and IRAC filters (IRAC2/3/4) following the methods outlined in \citep{Mondaletal2023, Sahaetal2024}. Using \texttt{SExtractor}\footnote{\url{https://sextractor.readthedocs.io/en/latest/Introduction.html}} \citep{r100}, we create segmentation maps and mask all segmented pixels to identify source-free regions. We then place 50 random 5$\times$5 pixel boxes in these regions, fitting a Gaussian distribution to the flux histogram of these apertures (see Appendix. \ref{back}). The mean and standard deviation of the Gaussian fit (see Figure \ref{back}), divided by 25 pixels, provide the background and noise per pixel. The mean background values per pixel are: AstroSat FUV F154W: \( 5.58 \times 10^{-6} \) count/sec, IRAC2: 0.017 MJy/Sr, IRAC3: 0.54 MJy/Sr, and IRAC4: 0.79 MJy/Sr. These values have been subtracted for subsequent analysis. 

PSF modeling is essential for correcting distortions from telescope optics and the atmosphere \citep{wolf2005, Liaudat_2023}. We model the PSF of a reasonably bright star using the Moffat function (Eq. [\ref{moffat}]) \citep{r96,r97}, as it better represents the extended wings of real stellar profiles compared to the Gaussian model (see Figure\ref{psf}). We use the \texttt{Astropy} Python library \citep{astropy} for the fitting, with the full-width at half maximum (FWHM) for SDSS ugriz bands being approximately 1.30" to 1.00" and FUV PSF FWHM of 1.35" from AstroSat \citep{r25,r30}. The IRAC PSF FWHM values are 1.66", 1.72", 1.88", and 1.98" for IRAC1/2/3/4, respectively \citep{r28}. For PSF matching, we convolve all images to match the broader PSF of the SPITZER IRAC4 band (FWHM = 1.99"). The PSF matching kernel(see \ref{psf_kernel}) is derived using the method in \citep{r49} and implemented through the \texttt{Photutils} Python library (see Figure. Top panel \ref{psf_kernel}). We apply this kernel to each filter's PSF and verify the matching effectiveness by comparing their growth curves (see Figure bottom panel \ref{psf_kernel}).

\begin{figure}
    \centering   
    \includegraphics[width=1.0\linewidth]{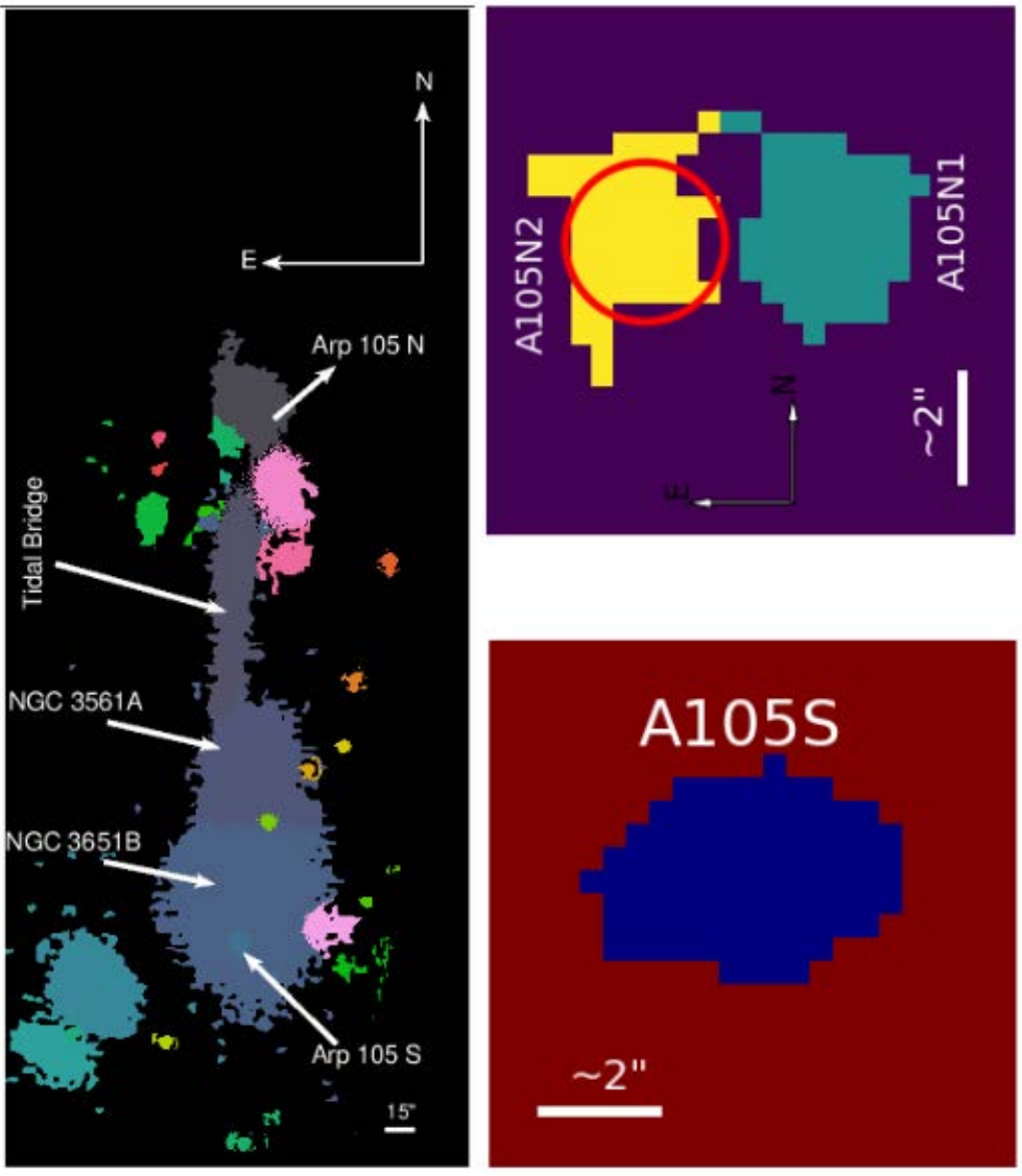}
   \caption{(Left panel): Segmentation map generated from the PSF-matched SDSS $r$-band image(pixel$\sim$ 0.6") of Arp 105 (see Figure.~\ref{seg}). The two interacting galaxies (NGC 3561A and NGC 3561B) and the extracted regions corresponding to A105N, the tidal bridge, and A105S are indicated with arrows and annotations. Orientation: north up, east left. Pixel coordinates are shown along the axes; the field of view is $400 \times 1000$ pixels. (Right panel): Segmentation map A105S(bottom) and 2 knots (A105N1, A105N1) in A105N(top)  in SDSS original r-bands. The red circle around N2 is the location of the SDSS BOSS spectrograph fiber with a diameter of 3".}
    \label{seg}
\end{figure}

Accurate photometric measurements require the identification and separation of individual sources to avoid contamination from blending with nearby objects. This method provides an improvement over traditional aperture photometry by allowing detailed profile fitting.
To isolate A105N, A105S, and the tidal bridge, we ran \texttt{SExtractor} \citep{r100} on the PSF-matched SDSS r-band image, producing the segmentation map shown in Figure.~\ref{seg}. This map labels contiguous pixels of each detected object with a unique integer. Any disconnected pixels associated with the target sources were manually masked and reassigned as needed. The key \texttt{SExtractor} parameters are detailed in Table~\ref{sext_table}.

We extracted sources consistently across all bands by running SExtractor in \textit{dual-image} mode, adopting the PSF-matched SDSS $r$-band image as the reference for detection. The segmentation map derived from this band was used to measure fluxes in the other images.

Fluxes are susceptible to both foreground and internal dust extinction. \citep{r34} full-sky 100 $\mu m$ map provides us with the foreground dust reddening E(B-V)=$(0.0287\pm0.0003)$mag along a given line of sight of Arp 105. The values of the foreground extinction parameter for different bands are then derived using $ A_{\rm \lambda}=k_{\rm \lambda} E(B-V)$ for FUV 0.28 and 0.17,0.13,0.10,0.08,0.06 mag for SDSS u/g/r/i/z bands, respectively.  where $k_{\rm \lambda}$ is obtained following Calzetti extinction law \citep{r33} equations(3) and (4).

It was observed that A105N contains two clearly visible knots (N1 and N2) in the SDSS r-band image. To extract their photometric properties, we have used SExtractor on the original(non-PSF matched) SDSS r-band image, generating the segmentation map shown in Figure \ref{seg}. Using a detection threshold of DETECT\_THRESH=1.5 and a minimum area of 30 pixels, we have successfully separated the two knots, which possess equivalent circular areas of approximately 1.80 and 1.70 arcsec, respectively. For A105S, it is 2.20 arcsec. We then mapped these apertures to the F154W filter image to measure the fluxes from both knots. The same has been done for A105S. For A105N, the centers provided in the SExtractor output have coordinates RA, Dec (167.8017779, 28.7654232 ) and (167.8029885, 28.7655502) for N1 and N2, respectively. Their compact photometric properties are summarized in(see Table. \ref{knot_prop}.) 

To quantify the emission within the faint tidal bridge in the AstroSat FUV 154W image (see Figure.\ref{RGB}), we use a segmentation map of the bridge from the SDSS r-band projected onto the F154W image. This yields a total flux of 0.1055 count/s corresponding to 21.86$\pm$0.08 mag. The local background contribution was estimated by repositioning the same segmentation map across five nearby source-free regions, resulting in a mean background value of 0.0821 count/s. After the local background subtraction, the net flux turns out to be 0.0233 count/s for the source. Considering noise from the background as well as the source, the resulting Signal-to-Noise Ratio (SNR) yields a value of  SNR=5.56.

\begin{table}
\centering
\begin{tabular}{ |p{4cm}|p{3cm}|p{3cm}|  }
\hline
\multicolumn{2}{|c|}{Source Extractor Parameters} \\
\hline
Parameters & Value  \\
\hline
DETECT\_MINAREA  & 30   \\
DETECT\_THRESH   & 1.2   \\
ANALYSIS\_THRESH & 1.5   \\
FILTER\_NAME     & gauss\_2.0\_3x3.conv   \\
DEBLEND\_NTHRESH & 64   \\
DEBLEND\_MINCONT & 0.001    \\
CLEAN\_PARAM     & 1.0    \\
SEEING\_FWHM     & 1.99"  \\
\hline
\end{tabular}
\caption{First column represents parameter name and second column is the parameter values used for source extraction.  DETECT\_MINAREA is the minimum area detection corresponding to $\sim \pi\times(FWHM)^2$, DETECT\_THRESH is the level above the background level to detect as a source. FILTER\_NAME: Gaussian convolution kernel with a FWHM of 2.0 pixels, applied over a 3$\times$3 grid. DEBLEND\_NTHRESH: sets how many thresholding levels differentiate between overlapping objects. and DEBLEND\_MINCONT: control how sources get deblended. Seeing FWHM is 1.99" from IRAC4 PSF.}
\label{sext_table}
\end{table}

\begin{table}
\centering
\begin{tabular}{cccc}
Knots & F154W (AB) & SDSS-r(AB) & \\
\hline
A105N1  & 19.70  & 19.40& \\
A105N2  & 19.95  &19.66  &  \\
A105S & 18.03 & 18.20  & \\
\end{tabular}
\caption{Photometric properties of the identified compact sources. The multi-wavelength magnitude distribution compares the AstroSat FUV ($154\text{W}$ filter) and SDSS $r$-band corrected AB magnitudes. Individual knots are labeled by their respective survey identifiers}. 
\label{knot_prop}
\end{table}

\subsection{2D GALFIT modelling}
We use publicly available 2D \texttt{GALFIT} modeling software, provided by \cite{Peng}, to model our TDGs, which extracts structural components from galaxy images. It minimizes the difference between the original image and a parametric model convolved with the same band PSF. Since TDGs lack a well-defined morphology and are mostly irregular in shape, modeling them comes with certain limitations. Also, to date, there have not been many attempts to model TDGs in the literature. So, this study attempts to model A105N and A105S TDGs.  For modeling, we created 2D image cutouts (250$\times$250) and (40 $\times$40) pixels in size from the original SDSS r-band image for A105N and A105S, respectively. Initially, we provided the PSF image, the sigma image, along with the original r-band image. The centroid of the image was estimated using the \texttt{Photutils }library function $\mathrm{centroid\_com}$ \citep{r49}, which is determined from the image's first moment method.

\begin{figure*}
\includegraphics[width=1.\linewidth]{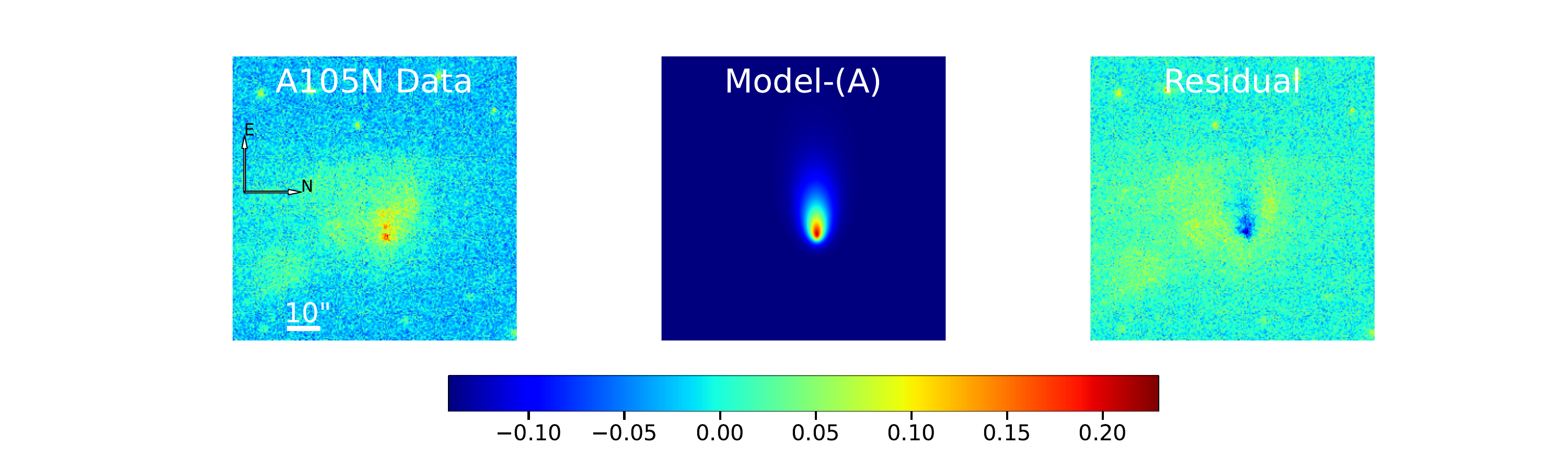}
\includegraphics[width=1.\linewidth]{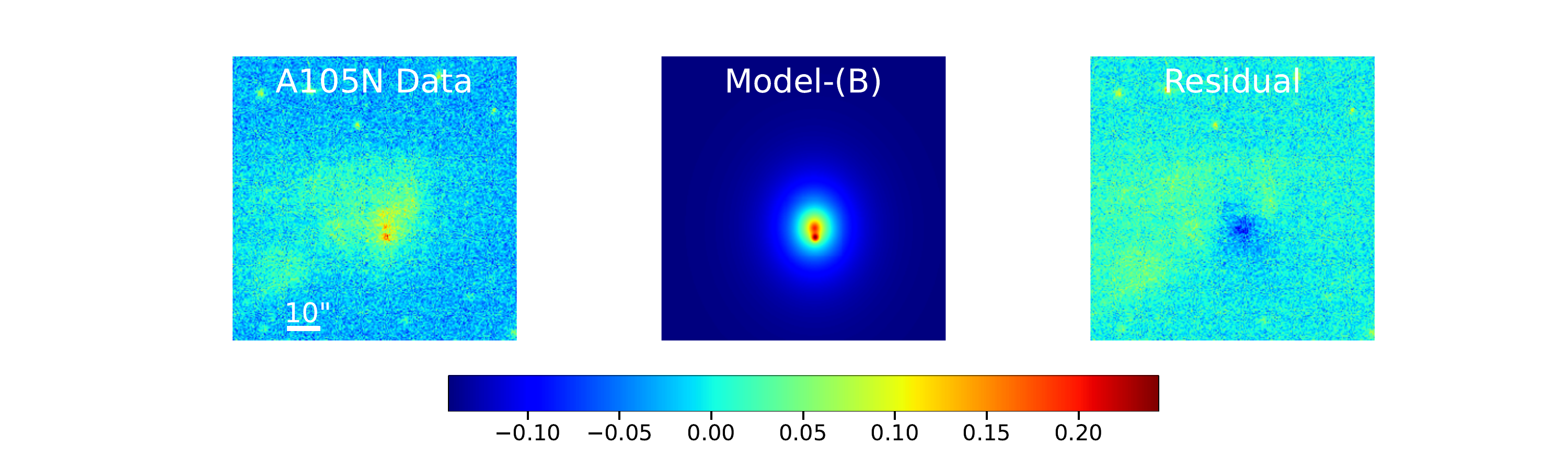}
\includegraphics[width=1.\linewidth]{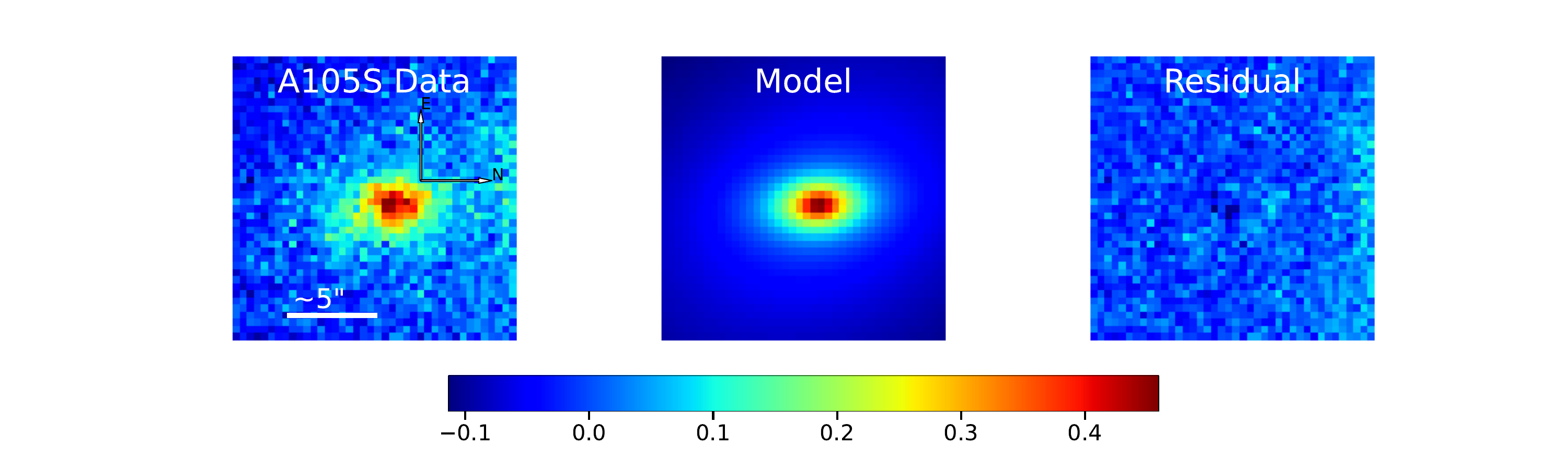}
\caption{ Two-dimensional \texttt{GALFIT }structural decomposition of the sample galaxies in the SDSS $r$-band. From top to bottom, the rows display the fitting results for A105N using Model (A), A105N using Model (B), and A105S, respectively. For each row, the panels from left to right show: (Left) the original SDSS $r$-band intensity map; (Middle) the best-fitting 2D Sérsic intensity profile; and (Right) the corresponding residual map (data minus model). Image cutouts have dimensions of $250 \times 250$ pixels for A105N  and $40 \times 40$ pixels for A105S, respectively. The colorbar indicates the flux in units of nanomaggies (nMgy). }
\label{105_2d_model}
\end{figure*}

A105N appears to be a very diffuse and asymmetric object (see Fig.\ref{RGB}), and hence, to account for asymmetry, we used the Fourier mode (m=1) with a Sersic disk (n=1) and fixed position angle(PA$\sim0^\circ$) in\texttt{ GALFIT} modeling (Model-A). We set the center pixel of A105N as fixed. The axis ratio $(0.77\pm0.029)$, Azimuthal Fourier mode 1 amplitude(A) =$0.75\pm 0.0073$ and Azimuthal Fourier mode(m=1) and phase($\phi=0.62\pm 1.42$ deg).  SDSS 1 pixel scale is 0.396 second, which gives the physical scale length of $1''=0.581$ kpc at z=0.0288, which gives an effective radius $R_\mathrm{e}(\rm A105N)=2.2\pm0.2$ kpc. 
For A105S, we modeled our galaxy with a Sersic disk light profile (n=1.27) and PA=$-87.05\pm4.36$ deg, and the effective radius is $R_ \mathrm{e}(\rm A105S)=1.31\pm0.2$\,kpc. It is observed that A105N contains 2 knots (N1, N2). So we also used the \texttt{GALFIT} Model-B (see middle row Figure.~\ref{105_2d_model}), to model these two knots separately, by implementing the Sersic light profiles to extract their geometric properties. Model-B is a better representation of the galaxy as it separates two bright knots in A105N. This gives an effective radius of $0.4\pm0.2$ with Sersic index n=$0.33\pm 1.14$ and $8.0\pm0.2$ kpc with n=$1.28\pm0.04$. For both components, the axis ratios (b/a) are 0.95$\pm$0.52 and $0.83 \pm0.01$ respectively. Note that our simplified Model B (see Figure.\ref{105_2d_model}), although a better representation of A105N, does not account for the diffuse light seen in the outer parts. Guided by the residual, our estimate suggests that approximately 25$\%$ of the total light is attributable to diffuse light in this galaxy. The diffuse light is distributed asymmetrically around the two knots in a similar fashion as in nearby starburst dwarf galaxies e.g., I zw 18, SBS 0335‒052E \citep{Papaderos2002, Herenz2023}. Furthermore, a disjointed structure of diffuse matter is also visible to the South-West side of the galaxy; this component was not included in Model B's estimate of diffuse light. The presence of all the diffuse light together does affect our estimate of the outer boundary, which extends up to $\sim 6\text{ kpc}$ to the south and $3.5\text{ kpc}$ to the north from the brightest knot, N1, and $15.7\text{ kpc}$ and $3.7\text{ kpc}$ in the east and west directions, respectively with a clear lopsidedness along the East-West direction. The boundary of the diffuse light corresponds to a surface brightness $24.60\text{ mag arcsec}^{-2}$.

\subsection{1D Surface Brightness Profile}
\begin{figure*}
\centering
\includegraphics[width=0.495\linewidth]{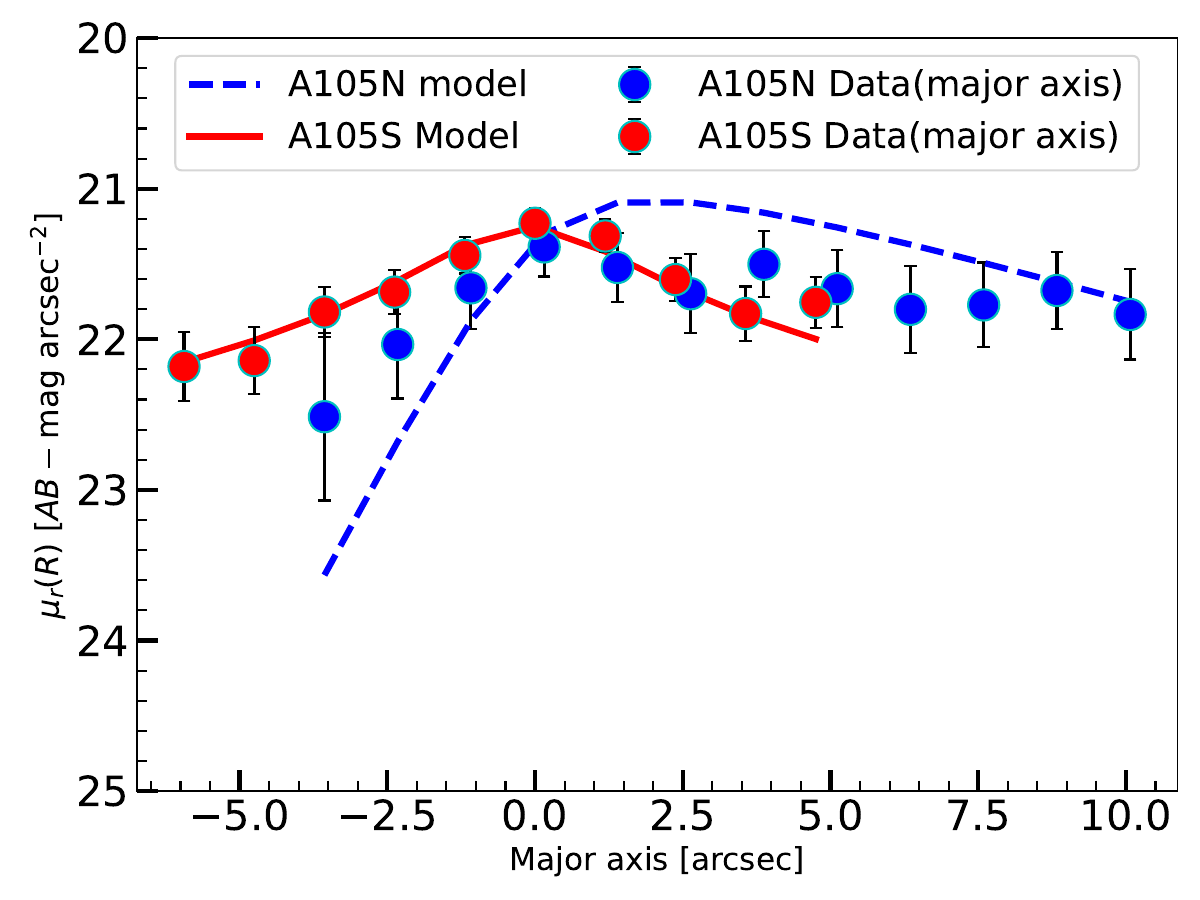}
\includegraphics[width=0.495\linewidth]{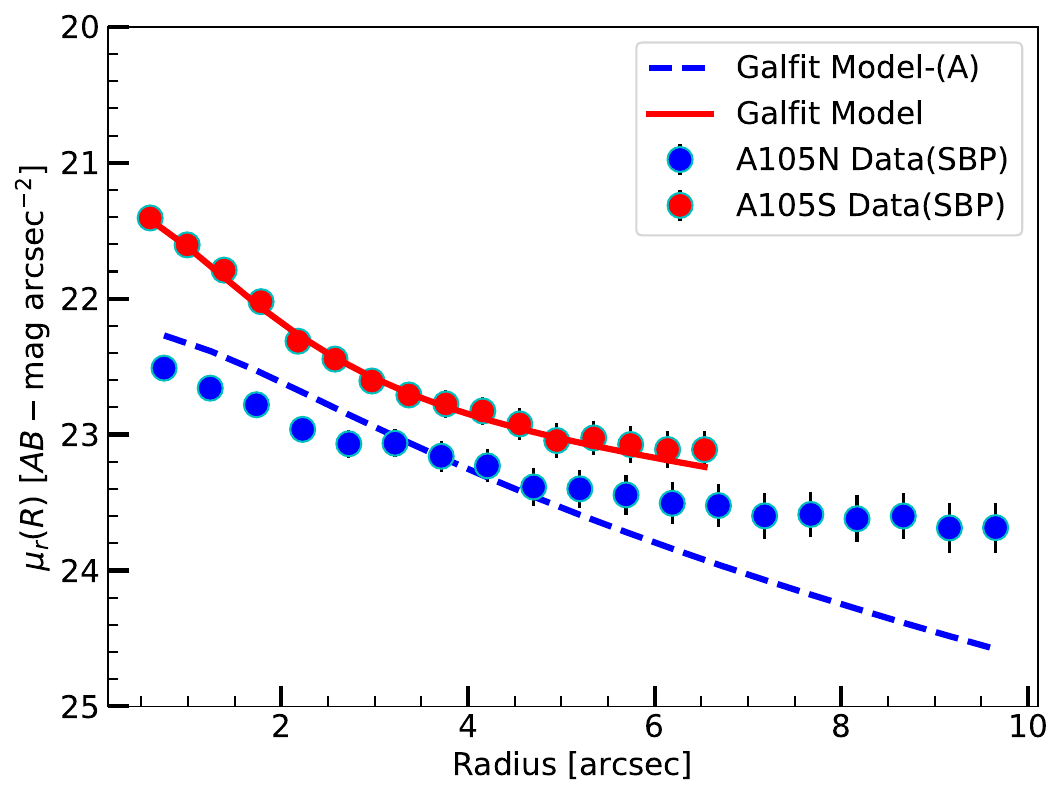}
\caption{The one-dimensional r-band surface brightness profiles of the two TDGs are shown. (Left panel:) The profiles are extracted along the major axis in the west–east direction for A105N (model-A) and in the south–north direction for A105S. (Right panel:) Radial profile of A105N (Model-A) and A105S created using annuli. For A105N, the blue points represent the observed data, while the blue dashed line shows the best-fitting model. For A105S, the red points denote the observed data, and the red solid line corresponds to the model. }
    \label{surface_brightness}
\end{figure*}

To model the surface brightness profiles, we extracted one-dimensional (1D) model profiles from the best-fitting two-dimensional (2D) flux distributions of the TDGs (see Figure \ref{surface_brightness}). Owing to the asymmetric and irregular morphology of A105N, firstly, we did not employ concentric circular aperture photometry to derive its 1D surface brightness profile. Instead, we placed a rectangular slit across A105N along the west–east direction passing through the center and measured the surface brightness by collapsing the flux within the slit along the major axis. For A105S, we adopted the same approach by placing a rectangular slit along the south–north direction to derive its one-dimensional surface brightness profile. In Figure \ref{surface_brightness}, the top panel shows the surface brightness profile of A105N, while the bottom panel corresponds to that of A105S. In addition to the profile along the major axis, we also used circular and elliptical annulus photometry to obtain a radial surface brightness profile (SBP) of A105S and A105N model-A, respectively(see  Figure.~\ref{surface_brightness}).

\section{Spectral properties of \text{Arp\,105 N}}
\label{sec:dynamical}

\begin{figure*}
    \centering
    \includegraphics[width=18cm,height=10cm]{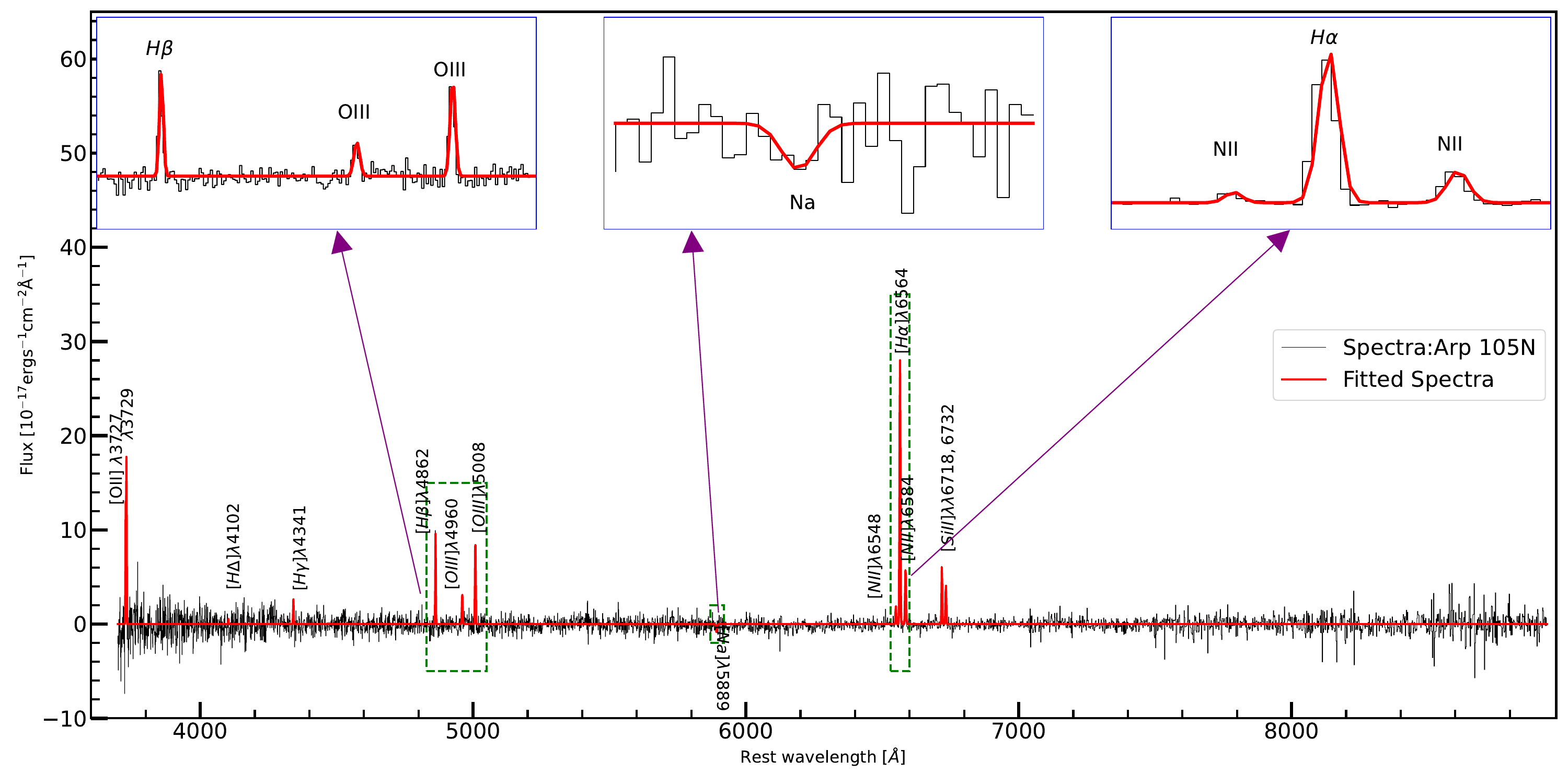}
    \caption{ SDSS spectrum of A105N in rest frame(z=0.029), X axis is rest frame wavelength in \AA\,  and Y-axsi is flux density in $10^{-17} \times \mathrm{erg\,s^{-1}\,cm^{-2}}$ \AA$^{-1}$. The black step line represents the continuum-subtracted spectrum. Gaussian fits to prominent emission lines are shown in red, with their corresponding rest-frame wavelengths (in angstroms) labeled above each line. The green dotted rectangles indicate regions that are shown as zoomed-in views within the blue inset boxes. Purple arrows connect these inset plots to their respective zoomed regions in the main spectrum.}
    \label{spectra}
\end{figure*}

We use the SDSS BOSS spectrum available for Arp~105N only; the spectra are centered around the coordinate RA: 167.80288 and Dec:28.76550 deg within a 3 arcsec diameter of fiber  (see Fig.~\ref{spectra}) to extract the relevant spectroscopic parameters. As the spectrum is not continuum-cleaned, the stellar continuum is modeled using Chebyshev polynomials implemented through the \texttt{Specutils} package in the \texttt{Astropy} \citep{r23} library, in which the spectrum is smoothed using a median filter to eliminate spikes. Emission-line fluxes are then measured by fitting Gaussian profiles to the continuum-subtracted spectrum using the \texttt{SciPy} library \citep{r94}. The resulting line fluxes are listed in Table~\ref{flux_table}.

\begin{table}
\centering
\begin{tabular}{|c|c|c|}
\hline
Spectral Lines[$\lambda_{\rm rest}($\AA)] & Flux ($\times10^{-17}$)  $\text{erg s}^{-1}\text{ cm}^{-2}$ \\

\hline
$H\alpha$[6564]  &  114.99$\pm3.33$     \\
\hline
$H\beta$[4862]  &  23.71$\pm$2.82      \\
\hline
 NII[6548]   & 6.74$\pm$1.86    \\
\hline
 NII[6584]  &  25.65$\pm$2.46    \\
\hline
 OIII[5008]   &  29.79$\pm$3.33   \\
\hline
  OIII[4960]  &  11.80$\pm$3.13   \\
  \hline

\end{tabular}
\caption{(First column): are the prominent spectral lines, square bracket numbers show $\lambda_{\rm rest}$. (Second column): represents the measured flux and error associated with the same. }
\label{flux_table}
\end{table}

The nebular color excess is estimated using the Balmer decrement method \citep{r32}, assuming Case~B recombination with an electron temperature of $T = 10^4$~K and an electron density of $n_\mathrm{e} = 100~\mathrm{cm^{-3}}$. Under these conditions, the intrinsic Balmer ratio is $(H\alpha/H\beta)_0 = 2.86$, and the nebular reddening is given by

\begin{equation}
E(B-V)_{\mathrm{neb}} = 1.97 \log \left[ \frac{(H\alpha/H\beta)_{\mathrm{obs}}}{2.86} \right].
\end{equation}

Here, $H\alpha$ and $H\beta$ are the observed line fluxes measured from the SDSS spectrum of Arp~105N (see Fig.~\ref{spectra} and Table~\ref{flux_table}). 
The $H\alpha/H\beta$ ratio is preferred due to the higher signal-to-noise of both lines and their sensitivity to dust extinction arising from their wide wavelength separation. Using the $H\alpha$  and 
$H\beta$ fluxes (see Table~\ref{flux_table}), we obtain nebular color excess as 
$E(B-V)_{\mathrm{neb}} = 0.45 \pm 0.05$~mag. To estimate the stellar continuum reddening, we adopt the Calzetti attenuation prescription \citep{r33}, using 
$E(B-V)_{\star} = 0.44\times\,E(B-V)_{\mathrm{neb}} = 0.19 \pm 0.02$~mag. This indicates that \text{Arp\,105 N} is a dusty, star-forming galaxy. These extinction values are applied throughout the subsequent analysis. 
The wavelength-dependent attenuation is computed as $A_{\rm \lambda} = k_{\rm \lambda} E(B-V)$, where attenuation coefficient $k_{\rm \lambda}$ is taken from the Calzetti extinction law \citep{r33}. 

\par
The emission-line equivalent widths of Arp~105N are 
$\mathrm{EW}(H\alpha) = (77.60 \pm 1.30)\,\text{\AA}$ and 
$\mathrm{EW}(H\beta) = (15.8 \pm 1.75)\,\text{\AA}$, indicating strong ongoing star formation. 
The large $\mathrm{EW}(H\alpha)$ implies a young stellar population dominated by massive OB stars producing significant ionizing radiation. 
Such high equivalent widths are characteristic of starburst regions and are commonly observed in tidal dwarf galaxies formed through galaxy interactions. 
In the context of the interacting \text{Arp\,105 N} system, these values support a recent ($\sim 5$--$30\,\mathrm{Myr}$, depending on SF history) episode of intense star formation likely triggered by tidal processes.

\par
We employ the strong line method to quantify the metallicity of \text{A105 N}, and fluxes were corrected using Balmer decrement and Calzetti attenuation relations \citep{r33}. We used attenuation coefficient as $k_{\rm OIII[5008]}$,$k_{\rm H\beta}$, $k_{\rm H\alpha}$ ,and  $k_{\rm NII[6584]}$ are 4.46,4.60,3.32, and 3.31 respectively. The oxygen abundance can be estimated using the O3N2 method \citep{r24}, following the relation:
\[
12 + \log[O/H] = 8.73 - 0.32 \times \text{O3N2} = 8.50 \pm 0.02
\]
where \(\text{O3N2} = \log \frac{[\mathrm{OIII]\lambda5008} / H_\beta}{\mathrm{[NII]\lambda6584} / H_\alpha} = 0.75 \pm 0.08\). The derived metallicity is \(Z \approx 2/3 \, Z_\odot\), indicating that \text{Arp\,105 N} is metal-rich, with \(Z_\odot\ = 0.0134\). This value is consistent with previous reports of a metallicity of \(12 + \log(O/H) = 8.6\) for \text{Arp\,105 N}, and \(8.40\) for its companion galaxy, A105S \citep{r71}. This suggests that tidal dwarf galaxies (TDGs) are metal-rich compared to classical dwarfs of similar size and luminosity, with metallicities approaching similar to disk of spiral galaxies.

\noindent Using the N2 method \citep{r24}, we also derive a metallicity of:
\[
12 + \log[O/H] = 8.90 + 0.57 \times \text{N2} = 8.53 \pm 0.02
\]
\noindent where \(\text{N2} = \log \frac{\rm [NII]\lambda6584}{H_{\alpha}} = -0.65 \pm 0.04\). Metallicities obtained from both the N2 and O3N2 methods are in quite good agreement with each other. We also determine the metallicities of the parent galaxies, NGC 3651B and NGC 3651A, to be $12 + \log[O/H]=8.82$ and $8.73$, respectively, indicating that the parent galaxies are more metal-enriched than \text{Arp\,105 N}.

\par
The strength of the 4000\,\AA\ break, $D_{\rm n}(4000)$, for \text{Arp\,105 N} has been directly calculated from the SDSS spectrum, obtaining $D_{\rm n}(4000)=0.98$, and independently from SED fitting (see section~\ref{sec:sed}), yielding a consistent value of $D_{\rm n}(4000)=0.97$. 
We adopt the definition of $D_{\rm n}(4000)$ from \citet{r64}, where the break is defined as the ratio of the average flux density in the wavelength ranges 4050--4250\,\AA\ and 3750--3950\,\AA\ \citep{r67}. 
The 4000\,\AA\ break arises primarily from metal absorption features in the atmospheres of cool, evolved stars and is therefore sensitive to the mean stellar population age \citep{r65,r66}. The measured value of $D_{\rm n}(4000) \simeq 1$ indicates an extremely weak or no break, consistent with a very young stellar population dominated by massive O- and B-type stars. This is characteristic of systems undergoing active star formation within the past $\sim5$--100\, Myr. 

\section{Spectral energy distribution modeling}
\label{sec:sed}
In this section, we estimate the stellar and dust masses of the two TDGs and, for the first time, estimate the stellar mass of the tidal bridge. We determine stellar masses using spectral energy distribution (SED) fitting with the publicly available SED fitting code, \texttt{CIGALE} \footnote{\url{https://cigale.lam.fr/}}\citep{r40}.\texttt{ CIGALE} models galaxy emission from the far-ultraviolet to the infrared using an energy-balance approach, combining stellar population synthesis, dust attenuation, and dust re-emission. The physical parameters are inferred within a Bayesian framework by fitting the observed multi-wavelength photometry. We use flux measurements from $10$ filters: SDSS u, g, r, i, and z; Spitzer IRAC 1–4; and UVIT FUV, which together provide strong constraints on the stellar mass. The flux measurement procedure for each filter is described in Section~\ref{sec:dataanalysis}. The following \texttt{ CIGALE} modules are being employed to model the star formation history, stellar populations, dust attenuation, and dust emission.

\begin{itemize}
\item[1.] We implement the \texttt{sfh2exp} module to model the star formation history (SFH). Since tidal dwarf galaxies (TDGs) originate from parent galaxies, they contain both older stellar populations inherited from the parent system and younger populations formed from gas converting into new stars. The \texttt{sfh2exp} model combines two exponentially declining functions: the first represents the long-term stellar population, while the second accounts for the more recent starburst activity. A higher value of tau\_main indicates that SFR is constant for the main stellar population, and a lower value of tau\_burst indicates that the rapid decline of star formation rate(SFR) of the late starburst population.  

\item[2.] We consider the \texttt{bc03} module by Bruzual \& Charlot (2003) to model the stellar population along with Salpeter’s initial mass function (IMF) \citep{r69}. 

\item[3.] We employ \texttt{dustatt\_modified\_starburst} (Attenuation Laws) \citep{r33} to take into account of dust attenuation owing to stellar and nebular emission. The color excess, $E(B-V)_{\rm nebular}$, has been computed using the Balmer decrement method \citep{r32}.

\item[4.] The dust emission mechanism is modeled by adopting the \texttt{Dale 2104 } module \citep{r70}. 
\end{itemize}

The SED was fitted using various \texttt{Python CIGALE }modules, as listed in the (see Table.
\ref{tab:sed_fitting_parameters}). Since we have spectra for A105N, we fixed metallicity and attenuation based on the Balmer decrement. We also fixed the uv\_bump\_amplitude to 0. Additionally, we tested the sensitivity of the SED fitting ($M_{*}$) to the choice of IMFs as well as to the power-law slope $\delta$ used to modify the attenuation curve.  We tested for the Salpeter (IMF = 0) and attenuation optical slope $\delta$.
A positive $\delta$ steepens the curve, resulting in stronger UV attenuation, while a negative $\delta$ flattens the curve, leading to weaker UV attenuation. We used the segmentation-mapped (see Figure. ~\ref{seg}) fluxes of three components to fit their SEDs over a broad range of parameters simultaneously.

The best-fit stellar mass values were selected based on two criteria: (1) the best reduced chi-squared value ($\chi^2_\mathrm{reduced}$), and (2) the requirement that the SED be dominated by emission lines, as tidal dwarf galaxies (TDGs) are actively star-forming systems.
Our measurements of stellar mass ($M_*$) for the three components are as follows: for A105N, $M_{\rm*105N} = (5.75\pm1.44) \times 10^9\mathrm{M}_\odot$, $M_{\rm *,young105N}$=2.98$\times10^6M_\odot$, $M_{*,\rm old105N}$=5.75$\times10^9\rm M_\odot$  with the best-fit $\chi^2_\mathrm{reduced}=2.3$ and attenuation slope $\delta=-0.4$ ; for the tidal bridge, $M_{\rm*Br} = (6.83\pm1.40)\times10^9,\mathrm{M}_\odot$, $M_{\rm*,young,Br}$=$1.40\times10^6\rm M_\odot$, $M_{*,\rm oldBr}$=$6.83\times10^9\rm M_\odot$ with $\chi^2_\mathrm{reduced} = 3.5$ and $\delta=-1.0$ ; and for A105S, $M_{*105S} =(8.20\pm3.05)\times 10^8,\mathrm{M}_\odot$, $M_{\rm *,young105S}$=5.41$\times10^6\rm M_\odot$, $M_{*,\rm oldBr}$=8.14$\times10^8\rm M_\odot$ with $\chi^2_\mathrm{reduced} = 1.5$ and $\delta = 0.3$ with Salpeter IMF, and their corresponding dust masses is $M_{\rm dust}$: $1.00\times 10^8\rm M_\odot$, $3.50\times10^7\rm M_\odot$,and $0.89\times 10^8\rm M_\odot$ respectively.

Alongside the SED-derived stellar mass estimates, we also calculated stellar mass ($M_*$) using (g-r) color–mass-to-light ($M_*/L_{g}$) ratio calibration. Such color-based $M_*/L_{g}$ methods are widely adopted, with numerous calibrations available in the literature \citep{r41, r79, r80, r81}, as they incorporate the effects of stellar population properties, dust attenuation, and chemical enrichment. For this work, we employed the relation proposed by \citep{r41}, adopting a Salpeter IMF. The coefficients $a_g = -0.306$ and $b_g = 1.097$ were adopted for the color–mass relation. The $(g-r)$ color was derived using g- and r-band fluxes measured from the segmentation map. The stellar luminosity $L_g$ is expressed in solar units ($\rm{L_\odot}$), representing the galaxy's luminosity in the g-band.

\begin{equation}
\log \left(\frac{M_*}{L_\mathrm{g}}\right) = a_\mathrm{g} + b_\mathrm{g} (g - r)
\label{color-mass}
\end{equation}

The stellar masses $M_*$ for A105N, tidal bridge, and A105S  obtained from the above calibration (see Eqn.\ref{color-mass}) are $3.03\times10^9\rm M_\odot$,  $3.18\times10^9\rm M_\odot$, and,  $0.73\times10^9\rm M_\odot$ and their $M_*/L_{g}$ ratio is 1.32,2.0,1.30 respectively. So stellar mass $M_*$ measurements from both the SED and color-mass methods give a similar order of magnitude and are in quite good agreement with each other. So color-based mass-to-light ratio ($M_*/L_{\rm g}$) estimates exhibit strong agreement with those derived from spectral energy distribution (SED) fitting \citep{Roediger2015}. Utilizing a mock galaxy sample, these authors demonstrated that optical color-based masses align with SED-fitting results to within a scatter of $\sim 0.2\,\mathrm{dex}$. Furthermore, this consistency holds when applied to observed galaxies spanning a stellar mass range of $M_* \sim 10^8 - 10^{11} \, \mathrm{M}_\odot$.

\cite{Leroy2019} provided a NIR calibration for the stellar mass-to-light ratio ($\Upsilon_*^{3.4}\sim \Upsilon_*^{3.6}$) based on GALEX and WISE photometry using the Kroupa IMF. Based on this IMF, they adopted $\Upsilon_*^{3.4} = 0.35 ~ \rm M_{\odot} L_{\odot}^{-1}$ for star-forming  galaxies. They also reported that $\Upsilon_*^{3.4}$ varies systematically from $\sim 0.2$ to $\sim 0.5 ~ \rm M_{\odot} L_{\odot}^{-1}$ in the sense that the stellar mass-to-light ratio at $3.4\,\mu\text{m}$ ($\Upsilon_*^{3.4}$) decreases as the specific star formation rate ($\text{SFR}/M_*$). Along with NIR, we also used the IRAC [3.6]-[4.5]  color-mass relation proposed by \cite{Meidt2012ApJ}, assuming a Chabrier IMF, which adopted a color-independent mass-to-light ratio of $\Upsilon_{3.6} = 0.6$ and an absolute magnitude of the Sun at 3.6$\mu$m is 3.24 mag in the Vega AB magnitude system. By effectively bridging old, metal-rich and young, metal-poor populations, this uniform baseline maintains a systematic uncertainty of $\sim0.1\text{ dex}$—a precision comparable to existing alternatives. A comparative measurement (current study versus others) of these stellar masses is shown (see Table. ~\ref{all_M*}.)

\begin{table}
    \centering
    \begin{tabular}{ccccc}
    Stellar mass &IMFs & A105N  & Bridge & A105S  \\
        \hline
 $M_{\rm *SED} \rm [M_\odot] \times 10^9$ &  Salpeter &  5.75  & 6.83   &  0.82\\
        
        \hline
    $M_*{\rm (g-r)} \rm [M_\odot]\times 10^9$ &Salpeter  &3.03 & 3.18 & 0.73\\
      \hline
   $M_*{\rm (Leroy)}\times 10^9$ &Kroupa&4.23  &3.71 & 1.56\\
   \hline
   $M_*{(3.6-4.5)} \rm [M_\odot]\times 10^9$&Chabrier & 7.26 & 6.36  & 2.68 \\
     \hline
        
    \end{tabular}
    \caption{Stellar Mass Comparison: SED Fitting, $(g-r)$ and NIR Color, and IRAC ($[3.6]\mu\text{m} - [4.5]\mu\text{m}$) Photometry for tidal components: A105N, Bridge, and A105S respectively.}
    \label{all_M*}
\end{table} 

Our measured stellar masses are consistent with those in the literature. Numerical simulations also suggest that typical TDGs have $M_* \sim 10^7$–$10^8\rm M_{\odot}$ \citep{r52}.
Most TDGs have stellar masses below $2 \times 10^8\,\rm M_{\odot}$, with a median of $2.4 \times 10^8,\rm M_{\odot}$ \citep{r8}. This estimate is close to the median value of $1.9 \times 10^8\,\rm M_{\odot}$ found for 407 TDGs by \cite{r42}. However, recent studies, based on the COSMOS survey, find TDGs with stellar masses $7.5 <\log{M/\rm M_{\odot}} < 9.5$ and relatively compact, with a few TDGs having effective radii larger than 1.3 kpc \citep{Renetal2020}.

From SED fitting, we derive a young stellar mass for Arp~105N of $M_{\star,\mathrm{young}} = 2.98 \times 10^6\,\rm M_\odot$ and a FUV-based star formation rate averaged over 100\, Myr of SED
$\mathrm{SFR}_{\mathrm{FUV}} = 3.35\,\rm M_\odot\,\mathrm{yr^{-1}}$, implying a characteristic formation timescale of $\sim1$~Myr. 
Comparable timescales are obtained for Arp~105S ($\sim4$~Myr) and the tidal bridge ($\sim1$~Myr), indicating a coeval episode of recent star formation across the system. 
These results are consistent with statistical studies showing that the young stellar populations in tidal dwarf galaxies typically have ages $\sim$6 Myr \citep{Hancock_2009} and median ages of $\sim30$~Myr \citep{r42}.

\begin{table*}
\centering
\noindent
\begin{adjustbox}{max width=0.5\textwidth, valign=t, margin=0em 0em 0em 0em}
\begin{tabular}{|l|l|l|}
\hline
\textbf{Parameter} & & \textbf{Value} \\ 
\hline
\multicolumn{3}{|c|}{\textbf{sfh2exp (Star Forming History)}} \\ 
\hline
tau\_main [Myr] & & 500,1000,\textbf{3000},5000\\ 
tau\_burst [Myr] & & 5,10,\textbf{20},40,50,75,90,120\\ 
f\_burst & & 0.02,\textbf{0.04},0.07,0.10 \\ 
age [Myr] & & 5000,7000,9000,\textbf{11000},13000\\ 
burst\_age [Myr] & &10,20,50,70,\textbf{90} \\ 
\hline
\multicolumn{3}{|c|}{\textbf{bc03 (Stellar Populations)}} \\ 
\hline
imf & & \textbf{0:Salpeter}, 1: Chabrier \\ 
metallicity & & 0.02 \\ 
separation\_age [Myr] & & 10 \\ 
\hline
\multicolumn{3}{|c|}{\textbf{nebular (Nebular Emission)}} \\ 
\hline
logU & & -3.0 \\ 
f\_esc & & 0.0 \\ 
f\_dust & & 0.0\\ 
z\_gas & &\textbf{ 0.012},0.014\\ 
\hline
\multicolumn{3}{|c|}{\textbf{dustatt\_modified\_starburst (Attenuation Laws)}} \\ 
\hline
E\_BV\_lines & & 0.45\\ 
E\_BV\_factor & & 0.44 \\ 
uv\_bump\_wavelength [nm] & & 217.5 \\ 
uv\_bump\_width [nm] & & 35.0 \\ 
powerlaw\_slope & &-0.4 \\ 
Ext\_law\_emission\_lines & & 1 \\ 
$R_v$ & & 3.1 \\ 
\hline
\multicolumn{3}{|c|}{\textbf{dale2014 (Dust Emission)}} \\ 
\hline
qpah & & \textbf{0.47}, 1.12, 1.77, 2.50 \\ 
umin & & \textbf{0.100}, 0.120, 0.150 \\ 
alpha & & 1.0,1.1,\textbf{2.0} \\ 
gamma & & 0.1 \\ 
\hline
\multicolumn{3}{|c|}{\textbf{redshifting (Redshifting)}} \\ 
\hline
- & & 0 \\ 
\hline
\end{tabular}
\end{adjustbox}
\vspace{0.5em}
\caption{Summary of input parameters used for spectral energy distribution (SED) fitting for A105N for the best-fit SED. Bold values among multiple parameters indicate the best-fit value or preferred values selected for the analysis across each module.}
\label{tab:sed_fitting_parameters}
\end{table*}

\begin{figure*}
    \centering
    \begin{minipage}{0.60\linewidth}
        \centering
        \includegraphics[width=\linewidth]{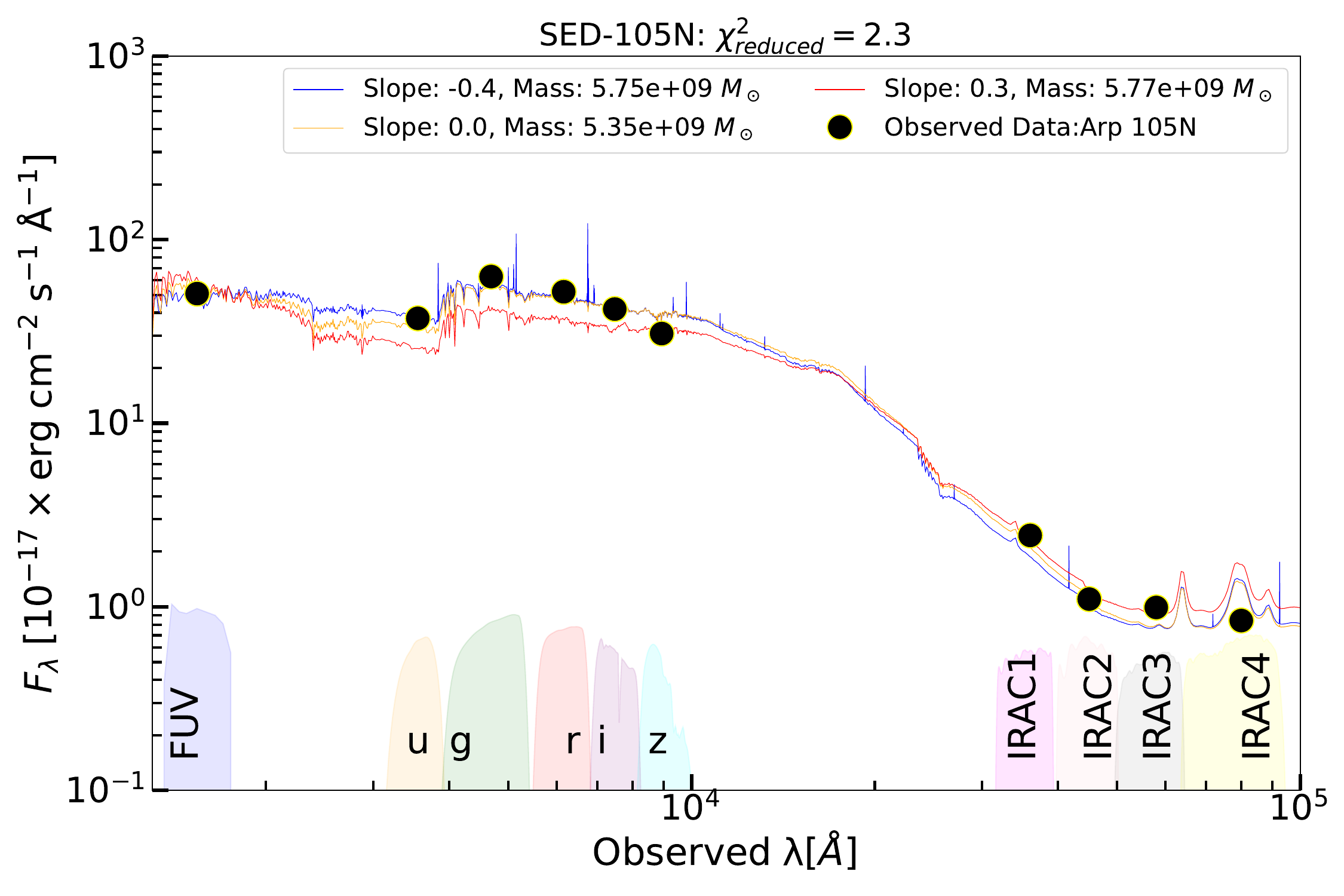}
    \end{minipage}
    \begin{minipage}{0.49\linewidth}
        \centering
    \end{minipage}
    
    \begin{minipage}{0.60\linewidth}
        \centering
        \includegraphics[width=\linewidth]{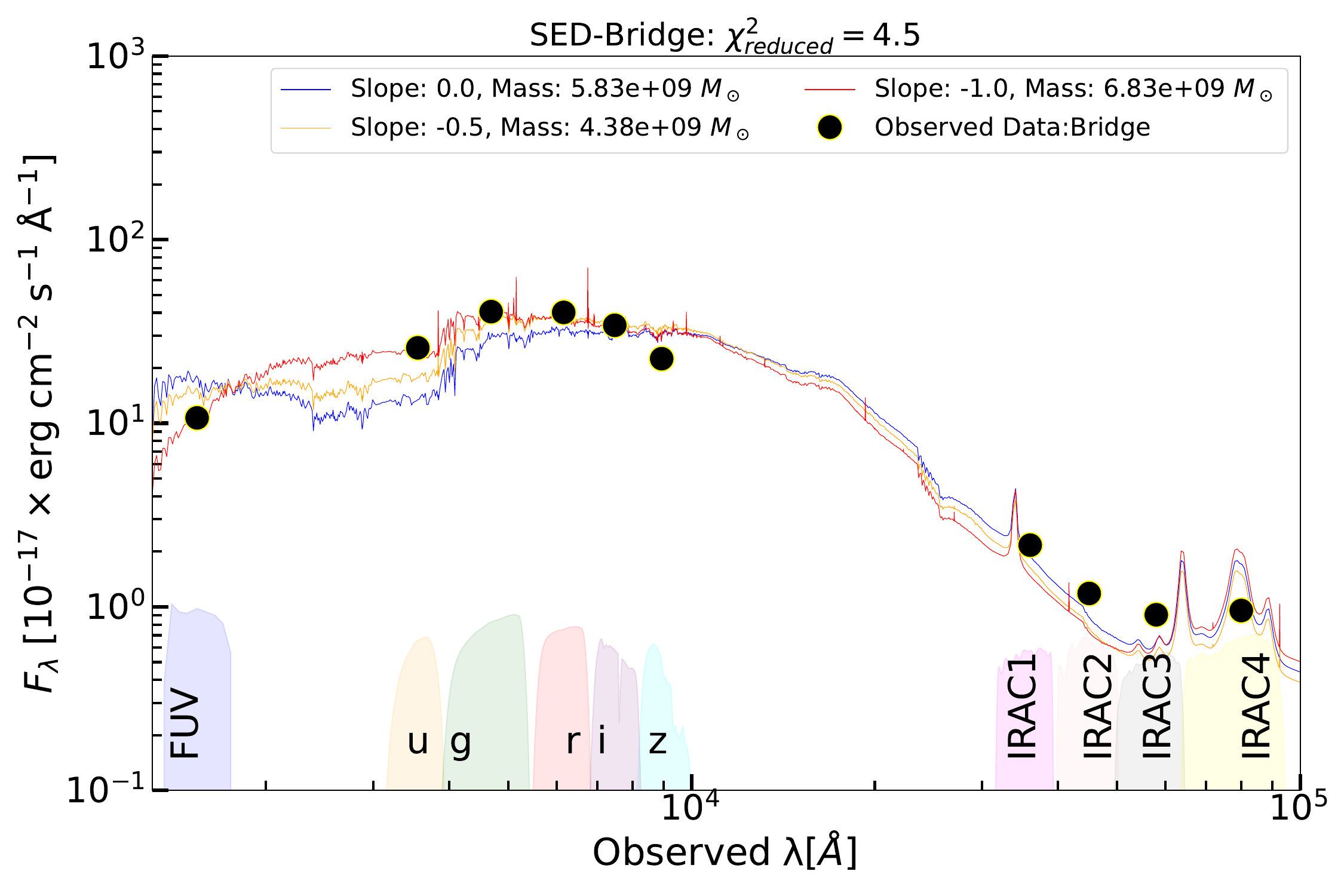}
       
    \end{minipage}
    \begin{minipage}{0.60\linewidth}
        \centering
    \end{minipage}

    \begin{minipage}{0.60\linewidth}
        \centering
        \includegraphics[width=\linewidth]{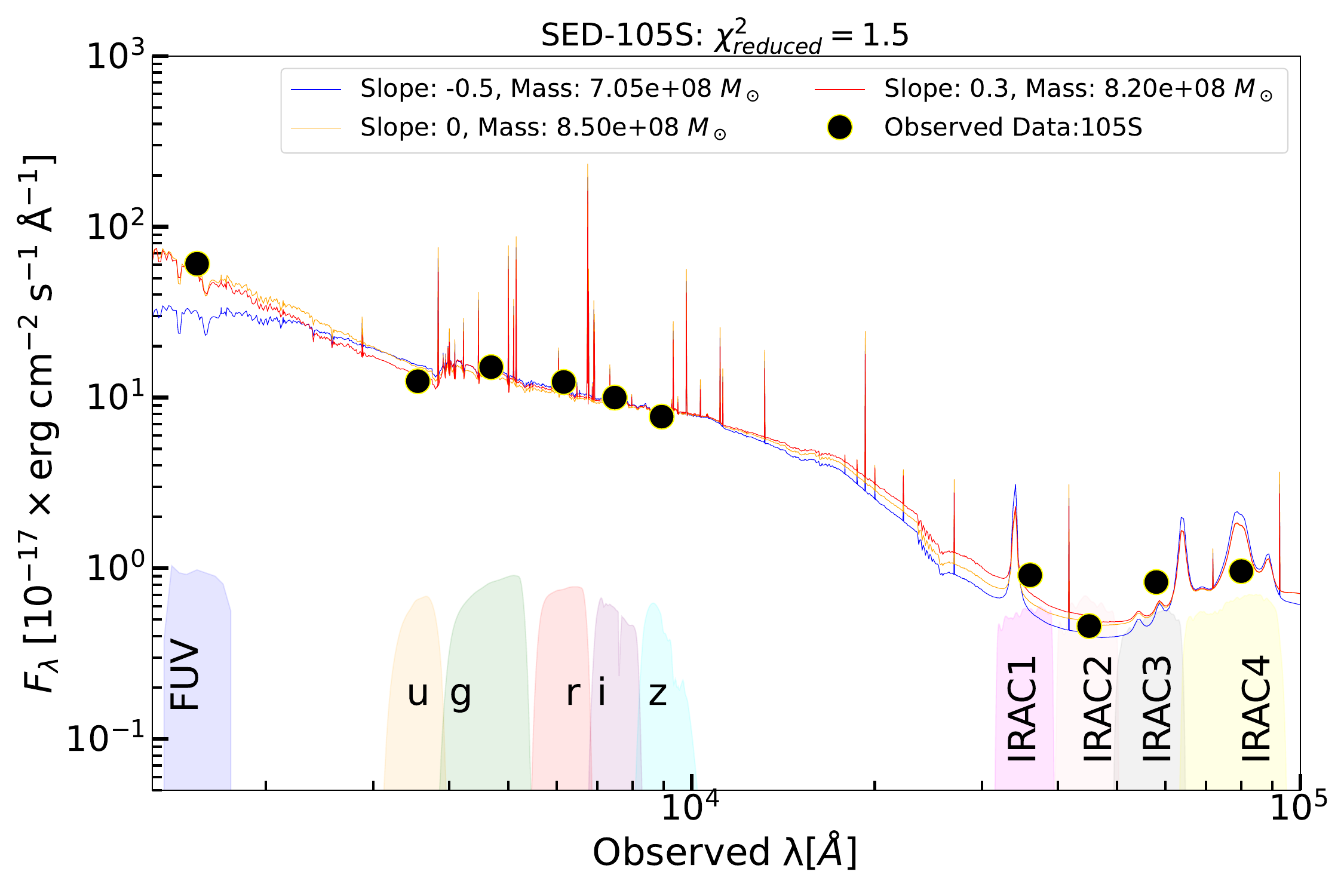}
    \end{minipage}
    \begin{minipage}{0.60\linewidth}
        \centering
    \end{minipage}
    \caption{Modeled SEDs for 3 tidal components with different colors with slope $\delta$. Top: A105N, Middle: Bridge, Bottom: A105S.  Black dots represent observed flux data. The blue curve is the best-fit SED for A105N, and the red curve is the best fit for A105S and the bridge tidal component. The best $\chi^2_{reduced}$ is marked on top of each figure. Inverted U-shaped curves are filter responses. Each SED is labeled with its corresponding stellar mass $M_*$.}
    \label{sed}
\end{figure*}

\section{sSFR--$M_*$ Relation}
\label{sSFR}

Galaxies sustain star formation by converting their gas reservoirs into stars, thereby building up their stellar mass ($M_*$). The evolutionary state of a galaxy can therefore be effectively characterized through the specific star formation rate (sSFR)--$M_*$ relation. In this work, we examine the location of the tidally formed systems in the sSFR--$M_*$ plane using star formation rates derived from $H\alpha$ emission and far-ultraviolet (FUV) fluxes, which probe complementary timescales of recent star formation.

The observed $H\alpha$ flux for A105N, measured from the SDSS fitted spectrum, is $(114.99 \pm 3.33) \times 10^{-17}\,\mathrm{erg\,cm^{-2}\,s^{-1}}$ (see Fig.~\ref{spectra}, Table~\ref{flux_table}). After correcting for both foreground and internal extinction using $k_{\mathrm{H}\alpha} = 3.32$, the extinction-corrected flux is $H\alpha[\rm corr] = 4.96 \times 10^{-15}\,\mathrm{erg\,cm^{-2}\,s^{-1}}$. The star formation rate $(0.053 \pm 0.003)\,\mathrm{M_\odot\,yr^{-1}}$ was estimated using the calibration(see Eqn. \ref{kroupa}) of \citep{Kennicutt_2012}, which assumes a Kroupa Initial Mass Function (IMF), and solar metallicity. 

\begin{equation}
\mathrm{SFR}[H \alpha] = 5.5 \times 10^{-42} L(H\alpha)\,(\mathrm{erg\,s^{-1}}) \ [\mathrm{M_\odot\,yr^{-1}}]
\label{kroupa}
\end{equation}

As we are using SFR calibration(See Eqn. \ref{salpeter}) with  Salpeter IMF \citep{r21} through this study, we get  $\mathrm{SFR}_{\mathrm{H}\alpha} = (0.076 \pm0.003)\,\mathrm{M_\odot\,yr^{-1}}$. Kroupa (Eqn. \ref{kroupa}) calibration gives lower SFR by a factor of $\sim1.5$.

\begin{equation}
\mathrm{SFR}[H \alpha] = 7.9 \times 10^{-42} L(H\alpha)\,(\mathrm{erg\,s^{-1}}) \ [\mathrm{M_\odot\,yr^{-1}}]
\label{salpeter}
\end{equation}

H$\alpha$ traces star formation over the past 5--10 Myr, probing the youngest stellar populations \citep{r59}. We also estimate star formation rates using extinction-corrected FUV fluxes from AstroSat/UVIT (F154W), applying both foreground and internal extinction corrections with $k_{\mathrm{FUV}} = 10.17$. FUV-derived star formation rates follow the calibration of \citet{r21}, which uses the Salpeter Initial Mass Function (IMF) with mass limits of 0.1 to
100$\rm M_\odot$ and stellar population models with solar
abundance. 

\begin{equation}
\mathrm{SFR}_{\rm FUV} = 1.4 \times 10^{-28} L_{\rm FUV}\,(\mathrm{erg\,s^{-1}\,Hz^{-1}}) \ [\mathrm{M_\odot\,yr^{-1}}]
\label{sfr_uv}
\end{equation}

which traces star formation averaged over the past $\sim$100 Myr. Using this relation, we obtain $\mathrm{SFR}_{\rm FUV} = 0.65 \pm 0.03$, $0.11 \pm 0.002$, and $0.77 \pm 0.008\,\mathrm{M_\odot\,yr^{-1}}$ for A105N, the tidal bridge, and A105S, respectively. The SDSS fiber diameter is 3 arcseconds, yielding a star formation rate of $\mathrm{SFR}_{\mathrm{H}\alpha} = (0.076 \pm 0.003)\,\mathrm{M_\odot\,yr^{-1}}$. The corresponding FUV flux within the same aperture is 0.006796 c/s; when multiplied by an extinction correction factor of 8.26 and an aperture correction factor of $\sim 1.55$ \citep{r25}, this results in $\mathrm{SFR}_{\mathrm{UV}} = (0.068 \pm0.030)\,\mathrm{M_\odot\,yr^{-1}}$. Consequently, both SFR measurements are in good agreement with one another.\\

\begin{figure*}
    \centering
    \includegraphics[width=0.8\linewidth]{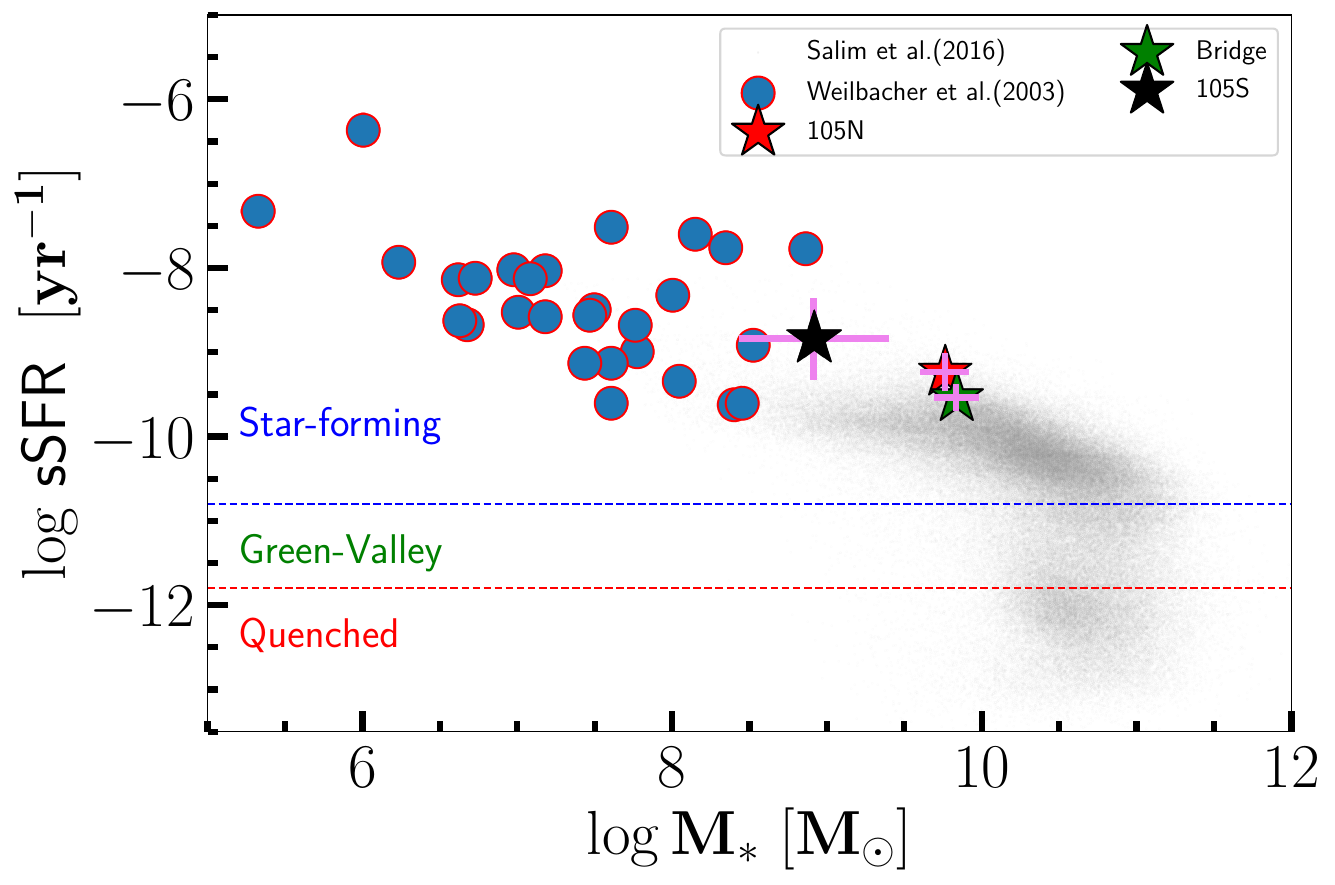}
    \caption{In this SED-derived Mass–sSFR diagram, gray data points are from \citep{Salim2016}. Blue circles represent data from \citep{Weilbacher2003}. The red, black, and green stars correspond to A105N, the tidal bridge, and A105S, respectively, from this study. The error bar is in magenta. The plot is divided into three regions based on specific star formation rate ($\log(\mathrm{sSFR})$): Star-forming region: $\log(\mathrm{sSFR}) \geq -10.8$, Green valley: $-11.8 < \log(\mathrm{sSFR}) < -10.8$, Quenched region: $\log(\mathrm{sSFR}) \leq -11.8$.}
    \label{m_ssfr}
\end{figure*}

For comparison, star formation rates derived from SED fitting over 10 and 100 Myr timescales are $(0.30, 3.35)$, $(0.14, 1.98)$, and $(0.55, 1.19)\,\mathrm{M_\odot\,yr^{-1}}$ for A105N, the bridge, and A105S, respectively. The corresponding 100 Myr-averaged specific star formation rates for A105N, the bridge, and A105S are $\log{\rm{sSFR}}$ $=-9.23$, $-9.53$, and $-8.83$ $\mathrm{{yr}^{-1}}$ from SED fitting, and $-9.94$, $-10.79$ $\mathrm{{yr}^{-1}}$, and $-9.02$~$\mathrm{{yr}^{-1}}$ from the AstroSat FUV measurements. All three tidally formed systems occupy the star-forming region of the sSFR--$M_*$ plane (see Fig.~\ref{m_ssfr}) and lie above the green valley \citep{r51}. A105N and A105S show significant FUV emission, consistent with ongoing star formation. The tidal bridge, although showing relatively weak FUV emission and lacking spectroscopic observations, still has a blue optical color in SDSS imaging (Fig.~\ref{RGB}). Its position above the green valley can be understood in terms of its low stellar mass: even modest absolute star formation rates produce high specific star formation rates (sSFR), keeping the system within the star-forming sequence. This is typical for low-mass, tidally generated systems, where the stellar mass is dominated by stars formed from material pulled from the parent galaxy, but ongoing star formation can still be triggered locally. Thus, even low-level star formation is sufficient to prevent the bridge from appearing quenched or in the green valley.

\section{Discussion and conclusions}
\label{sec:discuss}
Tidal dwarf galaxies (TDGs)/Non-classical dwarf galaxies and tidal bridges form from material expelled during interactions between gas-rich galaxies \citep{r60,r83,r14, Sengupta2013}. Unlike classical dwarfs, TDGs originate from recycled disk material and are relatively metal-rich compared to classical dwarfs of similar size and luminosity \citep{Duc1999, Smith2010,  Scott_2017, Recchi2015, Zaragoza_Cardiel_2024}. Their structural properties, including low concentration and high asymmetry, distinguish them from typical dwarfs\citep{r8}. Simulations and observational studies indicate that TDGs are largely devoid of dark matter, because they form from outer disk material, where dark matter density is generally low \citep{r17, Wetzstein2007, Bournaud2006, Bournaud_2009, Sengupta2013, Lelli2015, Sengupta2017, Ogiya_2021, Gray2023}. Owing to their enriched molecular gas content, TDGs are also expected to be detectable in CO emission \citep{Smith1994, Duc1999, Lisenfed2002, Braine_2000, kovakkuni2023}. These characteristics make TDGs compelling laboratories for exploring alternative pathways of galaxy formation and evolution.

In the context of evaluating the tidal interaction models within the Arp 105 system, the dynamical models provided by \cite{Holincheck2016} offer a critical framework for understanding its complex architecture. Their simulations reconstruct the orbit trajectories and disturbed morphologies of the system, identifying a mass ratio of $1.167 \pm 0.391$ and a dynamical interaction timescale ($t_{\rm min}$) of $278.9 \pm 30.9$ Myr since the closest approach. These results are vital for comparing these model parameters with multi-wavelength photometry, which allows for a robust cross-examination of evolutionary timelines. Specifically, AstroSat UVIT data reveal younger stellar populations of $\sim 100$ Myr (FUV) and $\sim 200$ Myr (NUV) GALEX \citep{Smith2010}. This alignment between the $\sim$280 Myr dynamical "clock" and the recent burst of star formation provides strong evidence for the tidal origin of the dwarf components, as brief, high-intensity star formation rates are primary indicators of maturing TDGs. Based on H-alpha kinematic analysis \citep{Bournaud2004}, it appears that
the northern Tidal Dwarf Galaxy (TDG) in Arp 105 might be a result of line-of-
sight projection rather than a distinct, bound entity. In contrast, the southern
star-forming Regions show evidence of being a self-gravitating, rotating body, suggesting it is a more structurally independent object.

It is interesting to note the extended, diffuse, and asymmetric distribution of material around A105N. This asymmetry is unlikely to be an intrinsic property of the galaxy itself, but is more plausibly related to its location within the interacting system. A105N lies near the tip of a long tidal tail, where part of the tidally stripped material may be falling back under the gravitational influence of the parent galaxy. Such returning tidal debris can lead to localized gas accumulation and asymmetries in the matter distribution. Importantly, the extended emission is detected not only in the r-band, but also in the far-UV, where the latter shows a somewhat clumpy morphology. Since the system exhibits ongoing star formation based on its emission-line properties, the far-UV emission is likely tracing young stellar populations formed within the last $\sim 100$ Myr. The spatial association between the FUV bright regions and the extended tidal material suggests that star formation is occurring locally within the tidal debris. A possible interpretation is that the fallback of tidal material creates locally compressive conditions within the gas, enhancing gas density and promoting star formation in these regions. This scenario is qualitatively consistent with numerical studies of interacting systems and TDG formation, where compressive tidal modes can trigger star formation inside dense tidal condensations. However, confirming the presence of a compressive tidal field would require higher-resolution kinematic information of the HI and ISM components, which is beyond the scope of the present work.

Classical or rather normal dwarf galaxies are known to be dark matter dominated \citep{Kormendy1987, Pryor1990}. Since TDGs are made from tidally stripped material from the outskirts of larger galaxies during tidal interactions that are insufficient to pull dark matter from those regions. In fact, dark matter density would be extremely low in the outskirts as is known from cosmological simulations. In other words, TDGs do not inherit any significant amounts of dark matter, as it cannot be captured by their weak gravitational potential \citep{Barnesetal1992, Bournaud_2009, Kroupa2012, Dabri_nd_Kroupa, Kroupa2015}.

We rely on the dynamical mass measurement from the VLA HI velocity map \citep{r50}. Based on these observations, an upper limit of $M_{\rm dyn} = 2.4\times10^{10}\,\rm M_\odot$ has been assigned to A105N. The baryonic mass of A105N is computed as $M_{\rm bary} = M_* + M_{\rm dust} + M_{\rm HI}$, where the stellar and dust masses are derived in Section~\ref{sec:sed}. For the atomic HI gas mass, we use the published HI measurements with $M_{\rm HI} = 6.6\times10^{9}$~M$_{\odot}$ \citep{r50,r63}. Since the dust mass is negligible compared to stellar mass, we estimate $M_{\rm bary} \approx  1.23\times10^{10}\,\rm M_\odot$; with a ratio $M_{\rm dyn}/M_{\rm bary} \approx 1.95$. Similarly, for A105S, dynamical mass measurement using the rotation curve determined in that object from H$\alpha$ observations $\sim 10^9\rm M_\odot$ by \cite{r2} and HI mass $5\times10^8\rm M_ \odot$ \citep{r50}, we obtain $M_{\rm dyn}/M_{\rm bary} \approx 1.30$. These ratios indicate that A105N and A105S are likely to be dark-matter deficient, consistent with expectations for tidal dwarf galaxies \citep{r17,r53, Bournaud_2009}. We think this study would benefit from an in-depth analysis, but with the addition of an IFU observation. \\
  
\noindent The key results obtained from this study are listed below:

\begin{enumerate}

 \item[1] Based on SDSS and UVIT/FUV observations, two star-forming knots are identified in TDG A105N, providing clear evidence of ongoing collapse of pre-enriched tidal gas and in situ star formation.  

  \item[2] Two-dimensional GALFIT modeling of SDSS $r$-band images yields effective radii of $2.22\pm0.2$ kpc and $1.31\pm0.2$ kpc for A105N and A105S, respectively, consistent with the spatial scales of dwarf galaxies.

 \item[3] The gas-phase oxygen abundance of A105N, $12 + \log(\mathrm{O/H}) = 8.53$ ($\sim 2/3\,Z_\odot$), supports a tidal origin, with pre-enriched material stripped from the outer disk of the parent galaxy.

\item[4] Multiple star formation diagnostics - including strong H$\alpha$ emission, large Balmer-line equivalent widths, elevated specific star formation rates, young SED-derived ages (5 - 10 Myr), UV emission tracing stellar populations younger than $\sim 100 - 200$ Myr, and weak 4000~\AA\ breaks with $D_n(4000)\sim 1$  ($<$ 1Gyr old) indicate the presence of a young stellar component superposed on an older population likely stripped from the progenitor disks.

\item[5] Our SED fitting yields stellar masses of $(5.75 \pm 1.44) \times 10^9\,\text{M}_\odot$ for A105N, $(6.83 \pm 1.40) \times 10^9\,\text{M}_\odot$ for the tidal bridge, and $(0.82 \pm 0.30) \times 10^9\,\text{M}_\odot$ for A105S. These values firmly place all three tidal components within the dwarf-galaxy regime. Furthermore, the inclusion of AstroSat FUV flux in the SED modeling provides a tight constraint on the stellar mass limit $\sim 10^{6}\rm M_\odot$ of the younger stellar population.

\item[6] The ratio $M_{\rm dyn}/M_{\rm bary} \sim 1.95$ and $\sim1.30$ for A105N and A105S, respectively, indicates a deficiency of dark matter in contrast to classical dwarf galaxies but consistent with theoretical expectations for tidal dwarf systems.
\end{enumerate}

\section*{Data Availability}  
The AstroSat UVIT FUV data underlying this article are not publicly available at the time of publication, but can be obtained upon reasonable request to the corresponding author.
The SDSS DR16 imaging (u/g/r/i/z bands) and spectroscopic data are publicly available in the Sloan Digital Sky Survey Data Release 16 archive at \url{https://www.sdss.org/dr16/data_access/}.
The Spitzer IRAC data (channels 1/2/3/4) are publicly available in the Spitzer Heritage Archive at \url{https://sha.ipac.caltech.edu/applications/Spitzer/SHA/}

\section*{Acknowledgement}

The authors are grateful to the anonymous referee for their helpful comments and suggestions, which improved the manuscript throughout the review process. This research work made use of various astronomical tools and \texttt{Python} libraries during this study. These are the following.

\begin{itemize}
    \item \texttt{SExtractor}
    \item \texttt{GALFIT}
    \item \texttt{CIGALE}
    \item \texttt{SAOImageDS9}
    \item \texttt{Numpy}
    \item \texttt{Matplotlib}
    \item \texttt{Scipy}
    \item \texttt{Astropy}
    \item \texttt{Photutils}
    \item \texttt{Specutils}
\end{itemize}




\bibliographystyle{mnras}
\bibliography{example} 

\section{Appendix}
\begin{appendix}
    \section{PSF Modeling and kernel matching} \label{psf_model}
    PSF was modeled with the Moffat function given below
\begin{equation}
I(r) = I_0 \left[ 1 + \left( \frac{r}{\alpha} \right)^2 \right]^{-\beta}
\label{moffat}
\end{equation}
\begin{itemize}
    \item \( I(r) \) is the intensity at radius \( r \),
    \item \( I_0 \) is the peak intensity
    \item \( \alpha \) is the scale parameter related with  width
    \item \( \beta \) is the power-law index that controls how steeply the wings fall off
    
\end{itemize}
The full width at half maximum (FWHM) of the Moffat profile is given by. In the limiting case of $\beta \to \infty$, the Moffat profile turns into a Gaussian.

\begin{equation}
    \text{FWHM} = 2 \alpha \sqrt{2^{1/\beta} - 1}
    \label{fwhm}
\end{equation}

\begin{figure}
        \centering
         \includegraphics[width=0.985\linewidth]{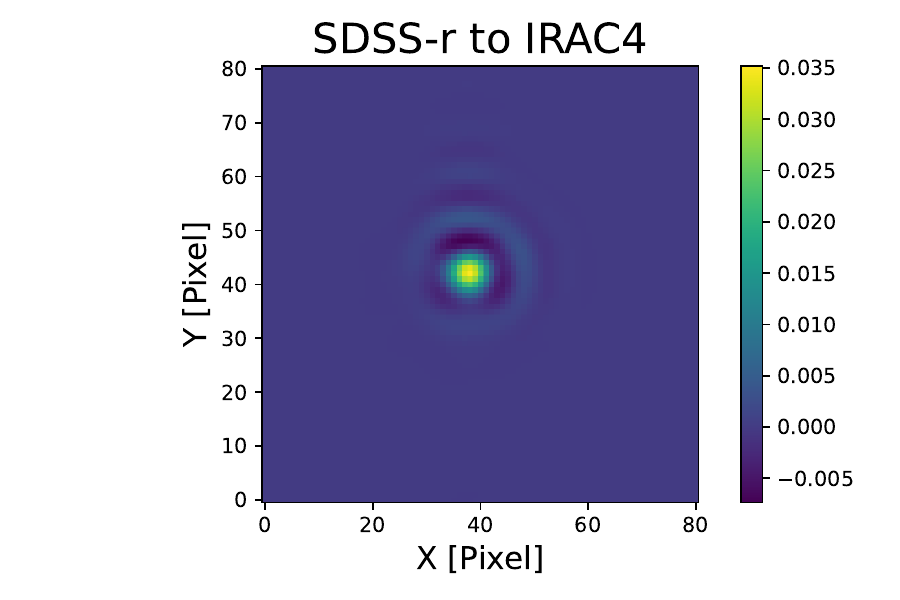}
          \includegraphics[width=0.95\linewidth]{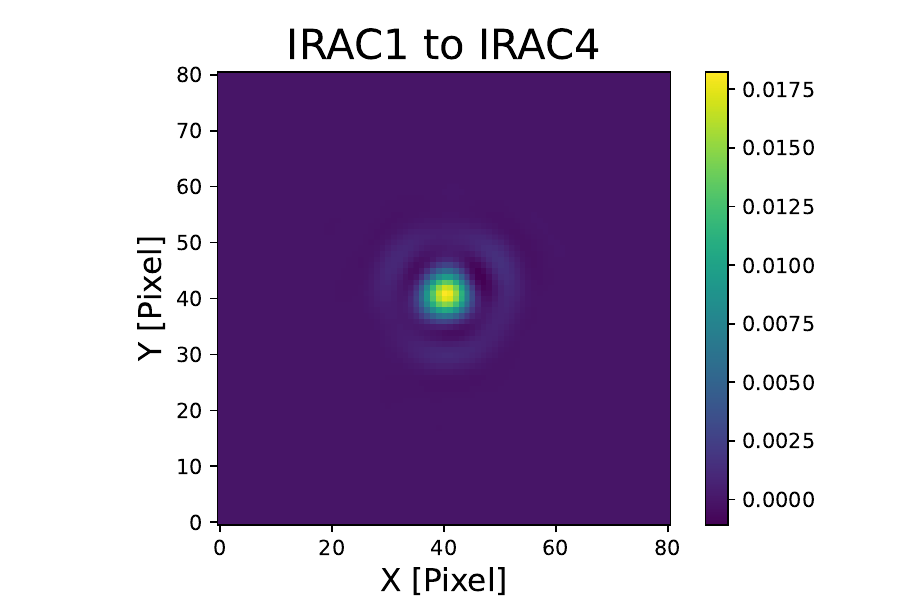}
        \includegraphics[width=0.75\linewidth]{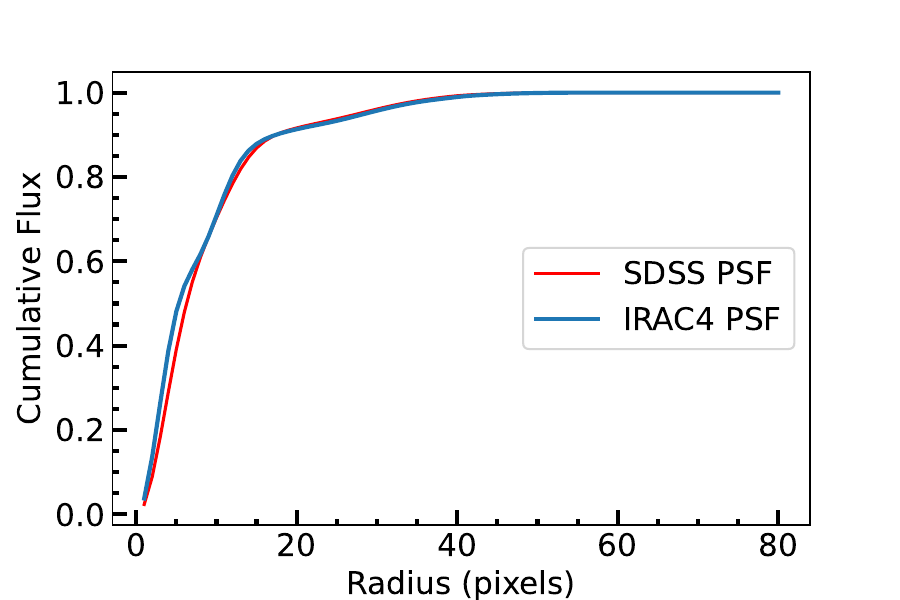}
        \caption{(Top and middle panel): Normalized convolution kernel required to match the SDSS r-band point spread function (PSF) and IRAC1 to that of the Spitzer/IRAC Channel 4 (8.0 $\mu$m) band. (Bottom panel): shows enclosed flux growth curves demonstrating the effectiveness of PSF homogenization, comparing the original SDSS r-band profile with the profile after convolution to the target IRAC4 PSF. }
    \label{psf_kernel}
\end{figure}

\section{Background Measurement}\label{back_plot}
We applied background subtraction to the raw (uncleaned) images in the IRAC Channels 2, 3, and 4, as well as the AstroSat FUV F154W band. After masking detected sources using the segmentation map, we placed 50 random 5$\times$5-pixel apertures across the image and constructed a histogram of their flux values. A Gaussian profile was then fitted to this distribution to determine the mean background level in each filter.
    \begin{figure*}
        \centering
        \includegraphics[width=0.4\linewidth]{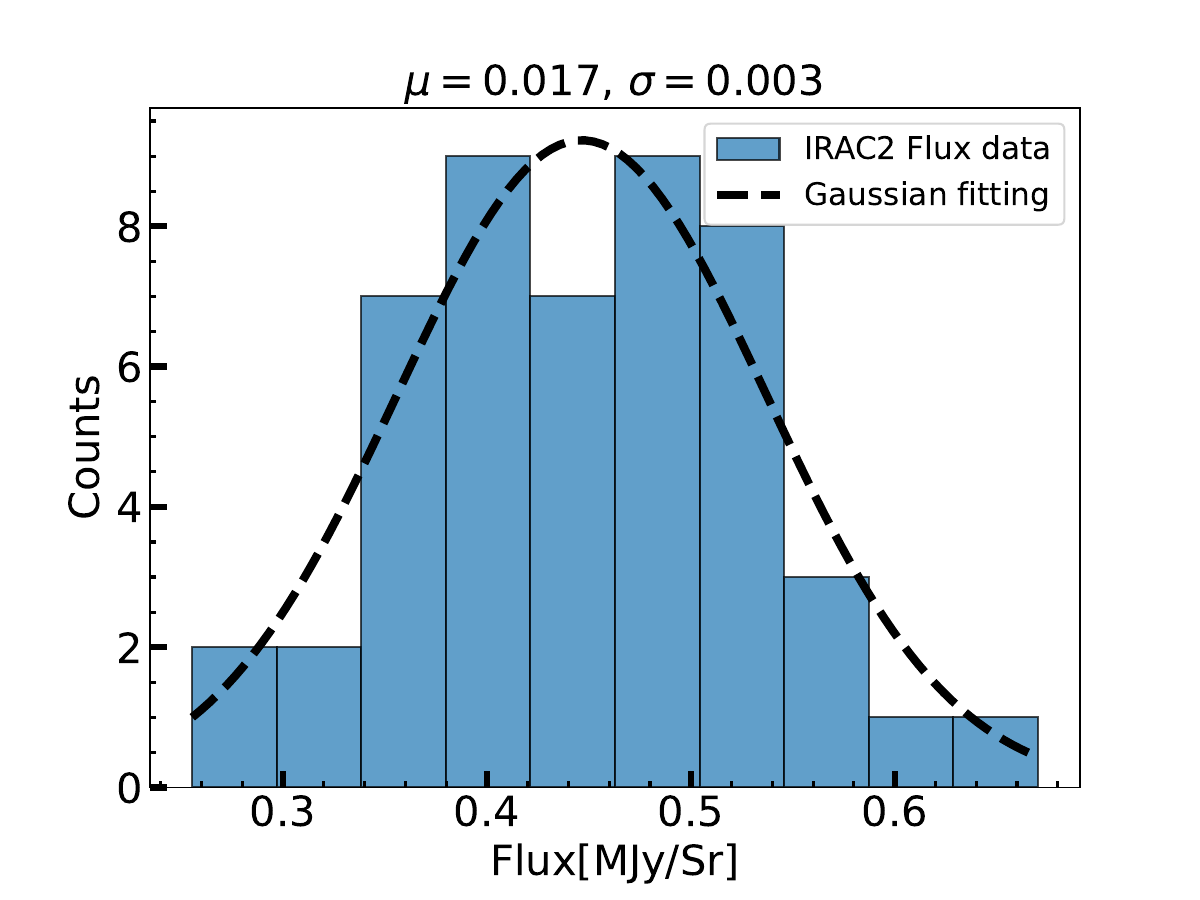}
        \includegraphics[width=0.4\linewidth]{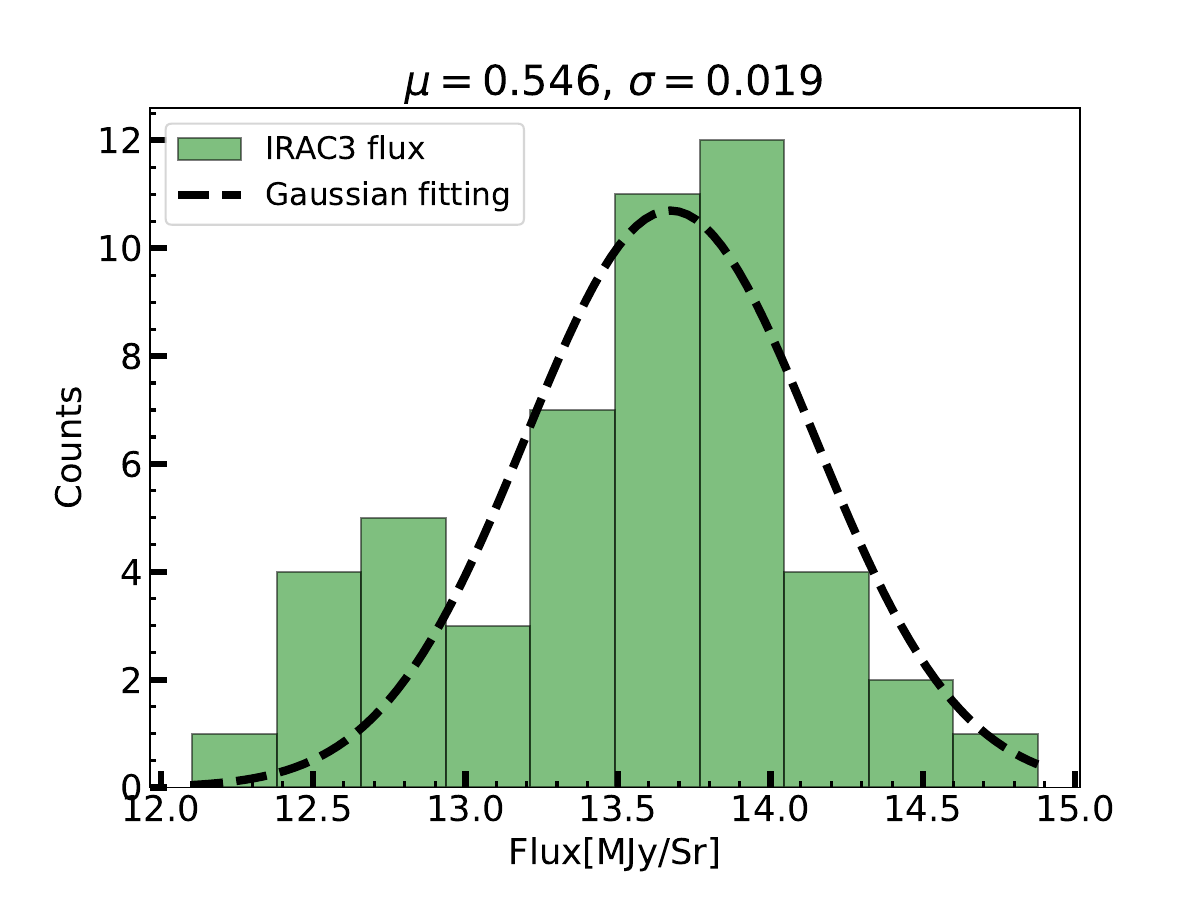}
        \includegraphics[width=0.4\linewidth]{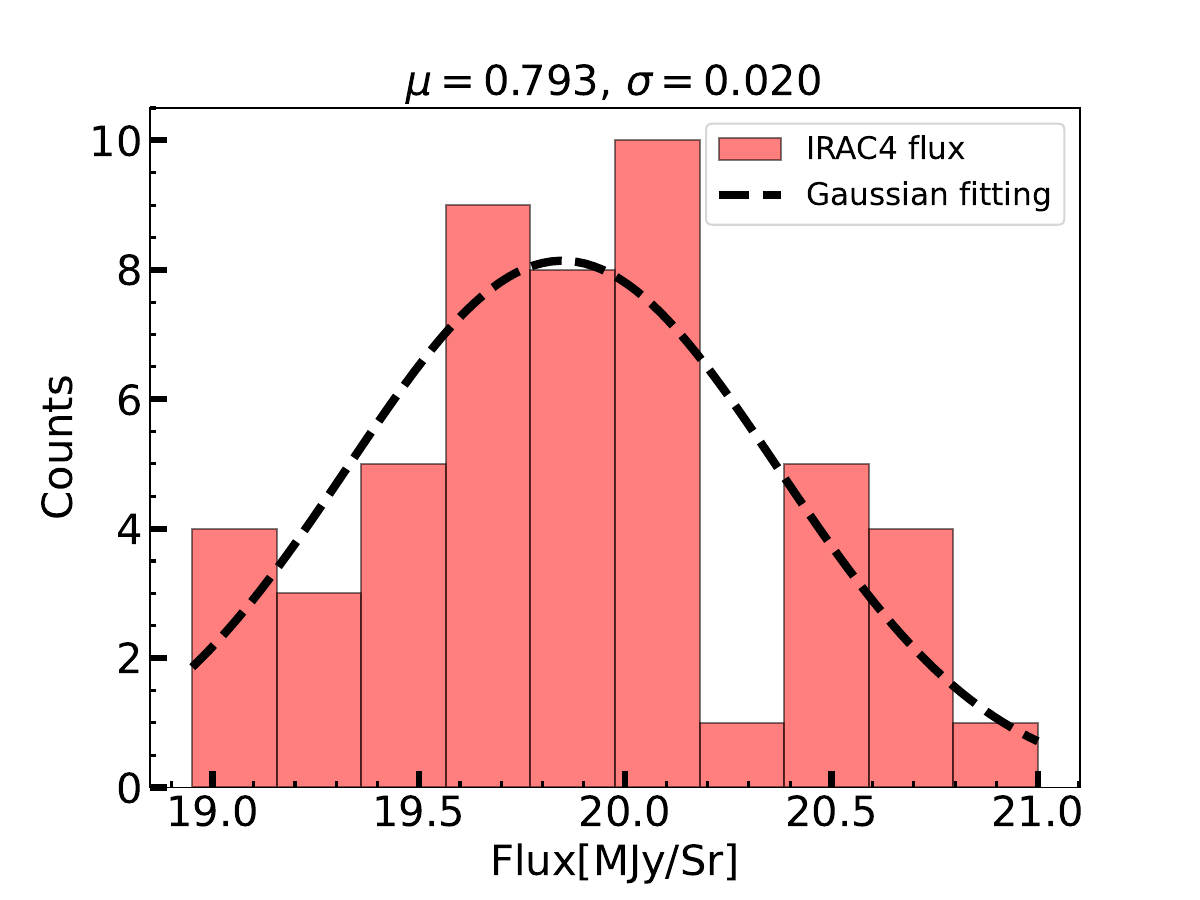}
        \includegraphics[width=0.4\linewidth]{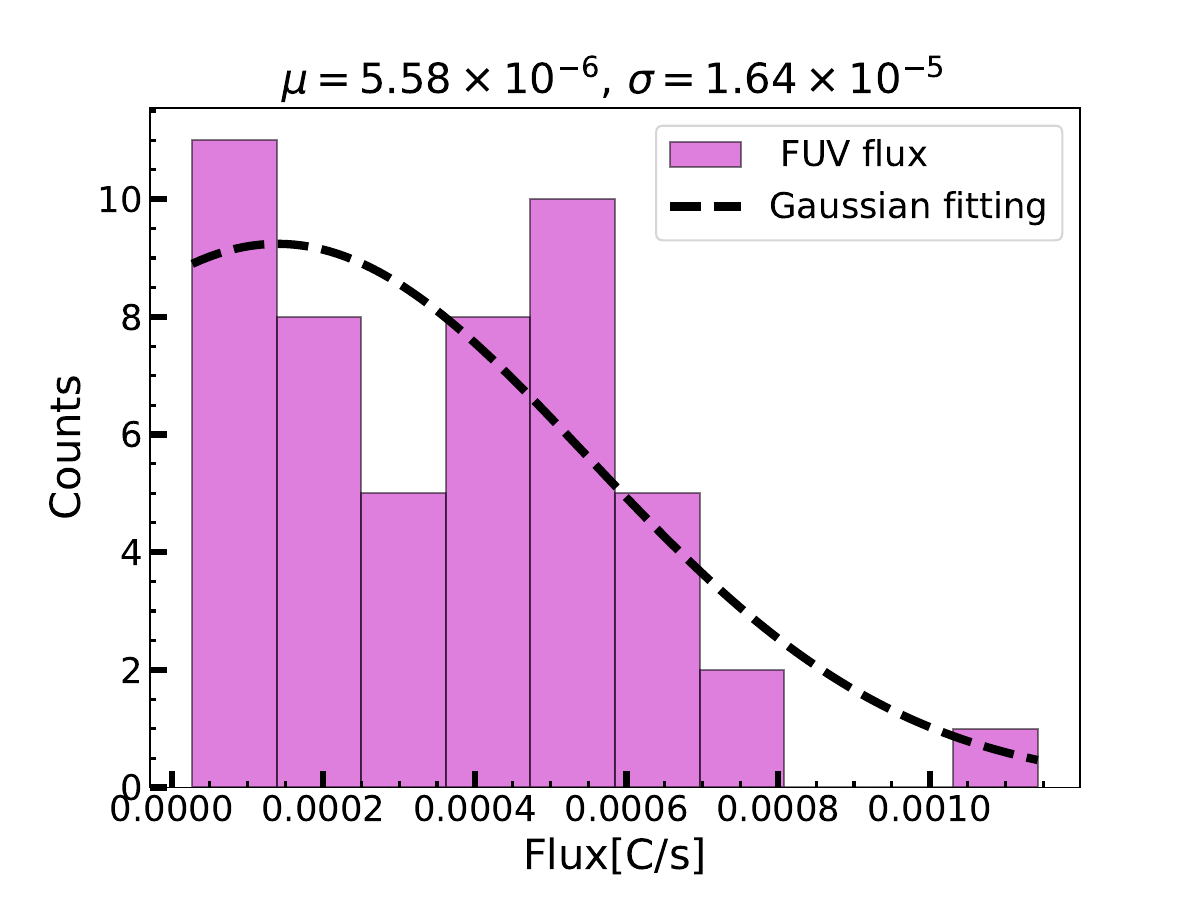}
        \caption{Different color histograms represent flux measurements in a 5x5 box size of IRAC2/3/4 and Astrosat FUV F FUV bands. The black dotted lines represent Gaussian fitting to the corresponding histogram. The X-axis is the flux bin, while the Y-axis represents the box counts.  Their best fit parameters: mean ($\mu$) and standard deviation ($\sigma$) per pixel values are quoted on top of each figure.}
        \label{back}
    \end{figure*}

\end{appendix}

\end{document}